\documentclass[10pt]{article}
\usepackage{amssymb, amsmath, cancel}
\usepackage{graphicx,psfrag,epsf}
\usepackage{enumerate}
\usepackage{natbib}
\usepackage{url} 

\newcommand{\blind}{0}

\usepackage{amsmath,amssymb}
\usepackage{amsthm}
\usepackage{multirow}
\usepackage{natbib}
\usepackage{bm}
\usepackage{hyperref}
\usepackage[all]{xy}
\usepackage{graphicx,caption, subcaption}
\usepackage{float}
\usepackage{color}

\newtheorem{thm}{Theorem}[section] 
\newtheorem{lemma}{Lemma}[section] 

\newtheorem{definition}{Definition}[section]
\newcommand{\bed}{\begin{definition}}
\newcommand{\eed}{\end{definition}}

\newtheorem{theorem}{Theorem}

\newcommand{\rom}[1]{\uppercase\expandafter{\romannumeral #1\relax}}

\usepackage{bbm}

\newcommand{\bitem}{\begin{itemize}}
\newcommand{\eitem}{\end{itemize}}

\newcommand{\goto}{\rightarrow}

\newcommand{\beq}{\begin{equation}}
\newcommand{\eeq}{\end{equation}}
\newcommand{\balign}{\begin{align}}
\newcommand{\ealign}{\end{align}}

\newcommand{\diag}{\mathrm{diag}}

\allowdisplaybreaks

\usepackage{enumitem}
\usepackage{makecell}

\usepackage{listings}
\begin{document}
	
\def\spacingset#1{\renewcommand{\baselinestretch}%
	{#1}\small\normalsize} \spacingset{1}


\if0\blind
{
	\title{\bf The SCORE normalization, especially for highly heterogeneous network and text data}
	\author{Zheng Tracy Ke and Jiashun Jin\\
		\vspace{0.4 em} 
	 Harvard University and Carnegie Mellon University} 
 \date{}

	\maketitle
} \fi

\if1\blind
{
	\bigskip
	\bigskip
	\bigskip
	\begin{center}
		{\LARGE\bf The SCORE normalization, especially for highly heterogeneous network and text data}
	\end{center}
	\medskip
} \fi

\begin{abstract}
SCORE was introduced as a spectral approach to network community 
detection. Since many networks have severe degree heterogeneity, 
the ordinary spectral clustering (OSC) approach to community detection may perform unsatisfactorily. 
SCORE alleviates the effect of degree heterogeneity  by introducing a new normalization idea in the spectral domain and makes 
OSC more effective. SCORE is easy to use and computationally fast. It adapts easily to new directions and 
sees an increasing interest in practice.  In this paper, we review the basics of SCORE,  the adaption of SCORE 
to network mixed membership estimation and topic modeling, and the application of SCORE in real data, including two datasets on the  
publications of statisticians. 
We also review the theoretical `ideology' underlying SCORE.   We show that in the spectral domain, 
SCORE converts a simplicial cone to a simplex, and provides a simple and direct link between the simplex and network memberships.  SCORE attains an exponential rate and a sharp phase transition in 
community detection, and achieves optimal rates in mixed membership estimation and topic modeling. 
\end{abstract}


\maketitle

\spacingset{1.45}

\tableofcontents

\spacingset{1}

\newpage

\section{Introduction}\label{sec:intro}  
In the era of Big Data,   large volumes of features and measurements are generated on a daily basis. Depending on the field, these could be deep sequencing data in cancer studies,  network data in social media,  and video data in self-driving car development. Many of such data sets are generated automatically and systematically. They are not custom-designed for any particular project one may have, and the data analysts may have little control on the data generating process.  
Also, many of such data sets are highly heterogeneous. The heterogeneity makes it hard to 
analyze the data, but in many cases, it is not directly related to the quantities of interest, 
and presents as an ancillary (or nuisance) effect.  
As examples, 
\begin{itemize} 
\itemsep0em
\item {\it Analysis of social networks}.  In many social networks, the degree of one node may be a few hundred  times larger than that of another. See Table \ref{tab:degree}, which is adapted from  \cite[Table 1]{SCORE+}.   
However, for many quantities of interest,  the degree heterogeneity is not directly related and presents as 
an ancillary or nuisance effect.  For example, suppose we want to use the co-authorship network to estimate the research areas or research interests of individual authors.  In this network,  an advisor and his/her advisee may have 
very different degrees (i.e., number of coauthors), but they may have very similar research interests and belong to the same research area. Therefore, the degree heterogeneity may only have an ancillary or nuisance effect over research areas or research interests.  
\item {\it  Topic learning for text data}.  In large  text documents, the frequencies with which different words appear are  
highly heterogeneous, where one word may appear a few hundred times more frequently than another.  
Similarly, for some quantities of interest, the frequency heterogeneity only has an ancillary or nuisance effect. 
For example, suppose a text document discusses one of the 
  three topics:  ``crime",  ``finance", and ``politics", but we do not know which. 
If we observe the word ``gunshot",    we are reasonably sure 
that the document is on ``crime", for ``gunshot" usually only appears 
when ``crime" (among the three topics) is being discussed.  
If we observe the word ``July", it is hard to decide which topic is being 
discussed, for no matter which topic is being discussed, 
the chance that ``July" appears is roughly the same.  In this case, 
the appearing frequencies of``gunshot" and ``July"  may be quite different, but such a difference is not very helpful in  deciding the topics. 
\end{itemize} 
In such heterogeneous settings, data analysis is challenging, and classical approaches (e.g., MLE,   spectral clustering) may perform  unsatisfactorily (e.g.,  \cite{SCORE, zhao2012consistency}). 
SCORE and its  
variants 
are useful in both settings above. It is a normalization idea which aims to remove 
the heterogeneity effects by normalizing the principal components 
of the data matrix, and so makes it easier and more effective to estimate the quantities of interest.  
Under a heterogenous latent variable model,  
the SCORE normalization transforms a low-dimensional simplicial cone (in the spectral domain)  to a low-dimensional simplex   and gives rise to a convenient approach to estimating the quantities of interest.   

SCORE is easy to use and computationally fast. It adapts easily to a range of new 
problem areas and has received increasing interests in practice. 
In this paper, we review the basics of SCORE, adaption of SCORE to network membership 
estimation  and topic learning, and application 
of SCORE to the publication data of statisticians and the New York taxi data, 
among others.

\begin{table}[hbt!]  
\centering
\scalebox{.78}{
\begin{tabular}{lllll |   lllll |  lllll }
\hline
Dataset & $n$   & $d_{\min}$  & $d_{\max}$ & $\bar{d}$  & Dataset  & $n$  & $d_{\min}$ & $d_{\max}$ & $\bar{d}$ 
& Dataset & $n$    & $d_{\min}$  & $d_{\max}$ & $\bar{d}$    \\ 
\hline
Weblogs &  1222 &  1 & 351 & 27.35 &  Football & 110  & 7 & 13 & 10.36    & Polbooks &  92   &1&24&8.13 \\
Simmons & 1137 &      1 & 293 & 42.67  & Karate  &  34   &  1&17&4.59  & UKfaculty & 79 & 2 & 39& 13.97 \\ 
Caltech &   590 &  1&179&43.36 & Dolphins & 62   &1 &12 & 5.12  \\
\hline
\end{tabular}
}
\caption{Severe degree heterogeneity of some networks ($n$: number of nodes; $d_{min}$, $d_{max}$,  and $\bar{d}$:    minimum,   maximum, and average degrees, respectively). For many natural networks, the histogram of the degrees 
has a power-law tail, so the network has severe degree heterogeneity.}  \label{tab:degree} 
\end{table}

\subsection{The basics on SCORE and the simplex structure} \label{subsec:SCOREbasics} 
SCORE stands for {\it Spectral Clustering On Ratios-of-Eigenvectors}.  It was first introduced by
\cite{SCORE} as a normalization idea for network community detection.  \cite{SCORE} motivated SCORE by 
the polblog network data \citep{polblog}. The data set was collected shortly after the 2004 presidential election. After light data preprocessing (e.g.,  \cite{zhao2012consistency}),  we have an undirected network with $n = 1222$ nodes, where each node is a web blog addressing US politics, and each edge is a (two-way) hyperlink between two blogs. The network is believed to have two communities: democrat and republican, and each node is manually labeled as a democrat or a republican (these labels are treated as ground truth; we use them for illustration below, but will not use them in our procedure).  Denote by $A$ the adjacency matrix of the network, where 
\begin{equation} \label{DefineA} 
A(i,j) = \left\{ 
\begin{array}{ll}
1, &\qquad \mbox{if there is an edge between nodes $i$ and $j$}, \\
0, &\qquad \mbox{otherwise}. \\ 
\end{array} 
\right. 
\end{equation}  
As we do not count self edges, $A(i,j) = 0$ if $i = j$. 
For each node, whether it is a democrat or a republican is unknown, 
and the interest is to find out this information using the adjacency matrix $A$. This is known as the problem of community detection in network analysis, or the problem of (node) clustering in statistics.   

SCORE runs as follows.  For $1 \leq k \leq n$,  let $\hat{\lambda}_k$ be the $k$-th largest (in magnitude) eigenvalue of $A$,  
and let $\hat{\xi}_k$ be the corresponding eigenvector (which we call the $k$-th eigenvector).  Input 
$A$ and the number of communities $K$ (we take $K = 2$ for the polblog network).  
\begin{enumerate} 
\itemsep0em 
\item Obtain the first $K$ eigenvectors $\hat{\xi}_1, \hat{\xi}_2, \ldots, \hat{\xi}_K$ of $A$ (for later references, we write 
$\widehat{\Xi} = [\hat{\xi}_1, \hat{\xi}_2, \ldots, \hat{\xi}_K]$).  
\item Define the matrix of entry-wise ratios $\widehat{R}$ by $\widehat{R}(i, k) = \hat{\xi}_{k+1}(i) / \hat{\xi}_1(i)$, $1 \leq k \leq K-1, 1 \leq i \leq n$.   
\item For a threshold $T > 0$, let $\widehat{R}^*$ be the regularized version of $\widehat{R}$, where 
$\widehat{R}^*(i,k)=\mathrm{sgn}(\widehat{R}(i,k))\cdot\min\{T,\, |\widehat{R}(i,k)|\}$,  $1 \leq k \leq K-1, 1 \leq i \leq n$.  
\item Cluster all $n$ nodes into $K$ clusters by applying the classical $k$-means (with $K$ centroids) to the $n$ rows of $\widehat{R}^*$.  
\end{enumerate} 
In Step 2, we recommend $T =  \log(n)$ (and $T = 2 \log(n)$ if $n$ is relatively small).  
Alternatively, we may also set $T$ as the top $1\%$-quantile (say) of $\|\hat{r}_1\|,  \ldots,\|\hat{r}_n\|$. 
We assume the network is connected when we apply SCORE. 
Otherwise, we divide the network into connected components and then apply SCORE
to each component of interest. 
We call Steps 2-3 the {\it SCORE normalization}. 
If we skip these steps, then SCORE reduces to the 
Ordinary Spectral Clustering (OSC) algorithm,  where we cluster by directly applying the $k$-means to the $n$ rows of the 
matrix $\widehat{\Xi} = [\hat{\xi}_1, \hat{\xi}_2, \ldots, \hat{\xi}_K]$.  When the network is connected, 
the entries of $\hat{\xi}_1$ are either all strictly  positive or all  strictly negative \cite[Perron's Theorem]{HornJohnson}.  
  
\begin{figure}[htb!]
\centering
\begin{subfigure}[b]{.58\textwidth}
\centering
\includegraphics[width=1.5 in,height=1.5 in]{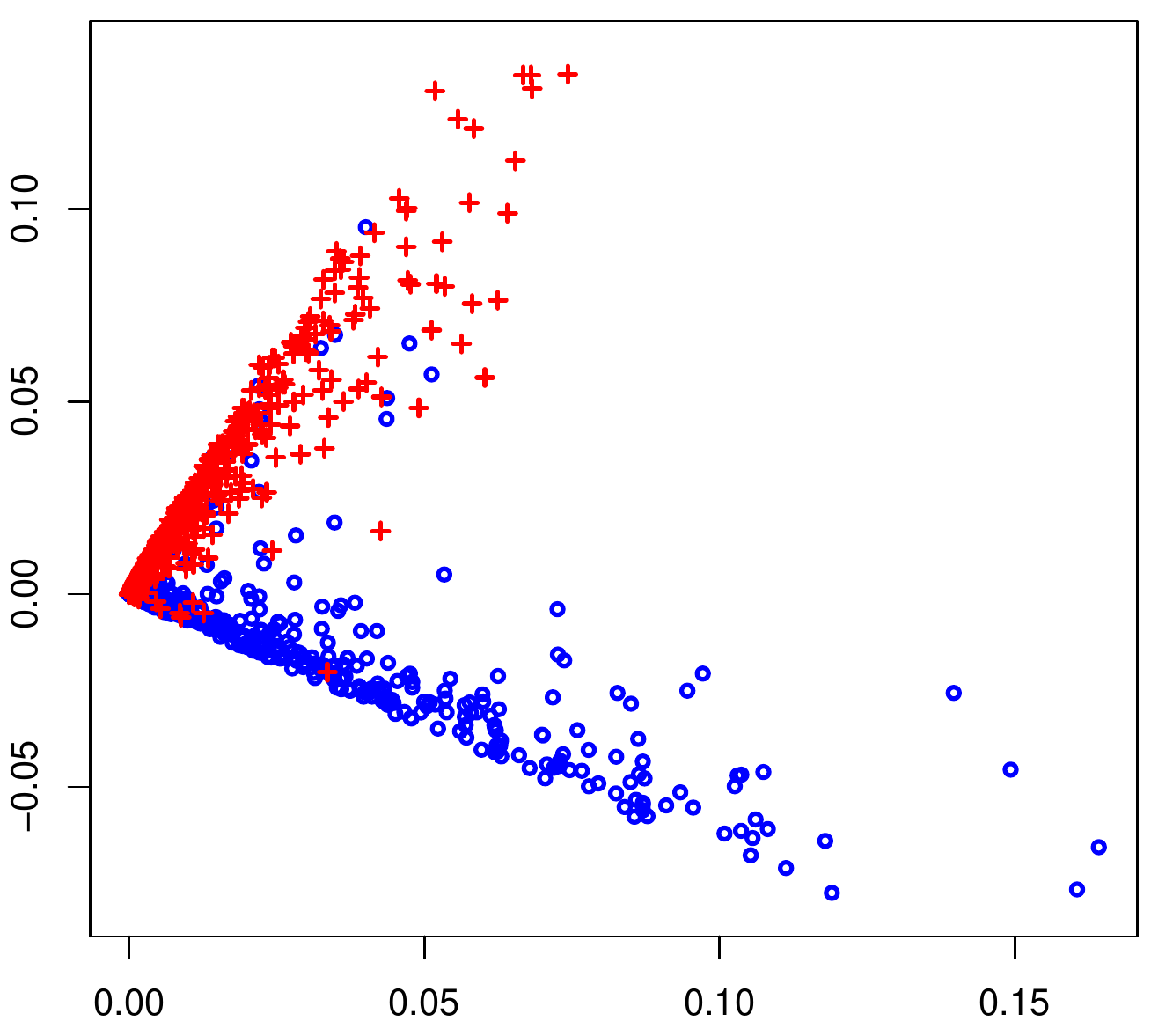}
\includegraphics[width=1.5 in,height=1.5 in]{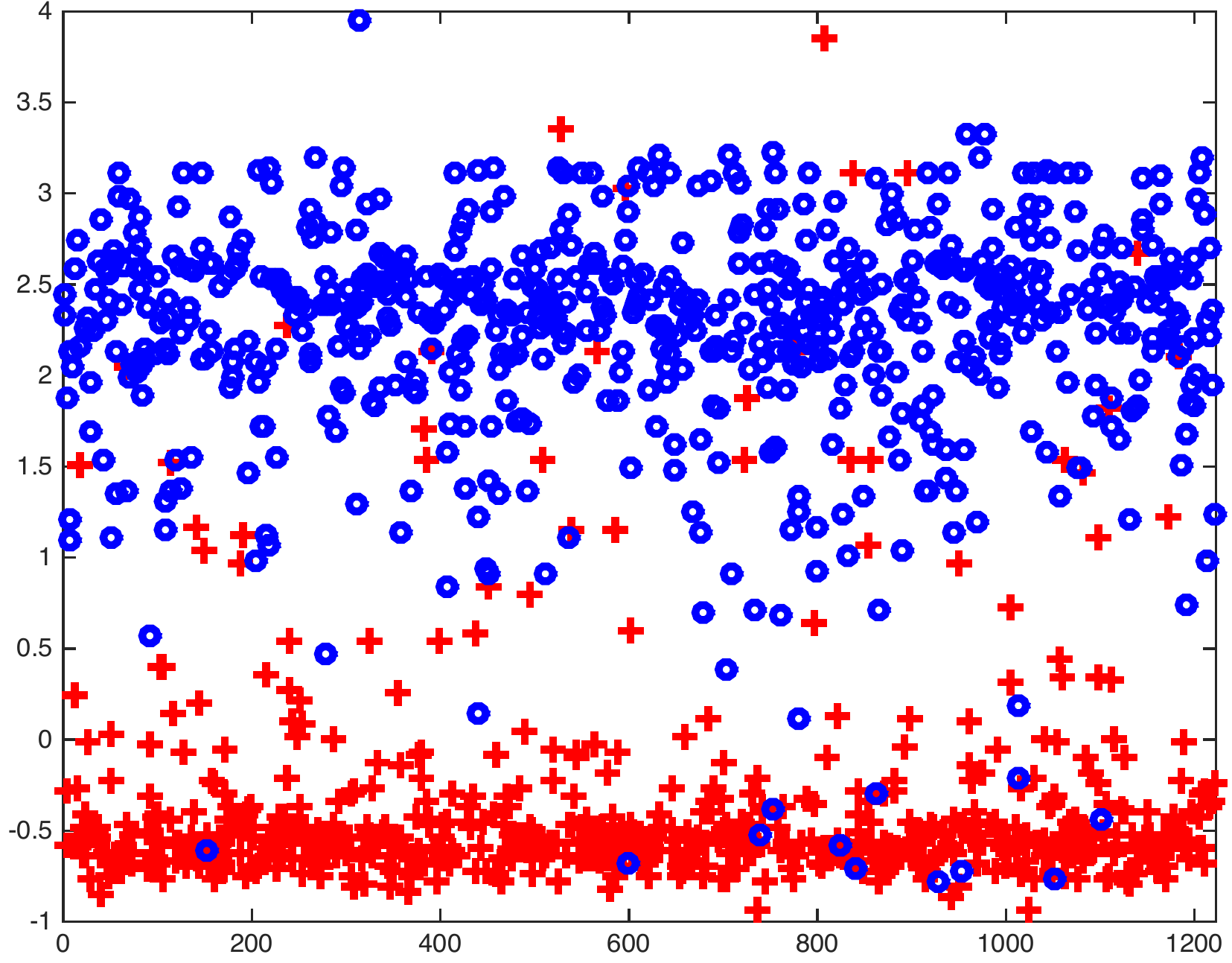}
\caption{The polblog data}
\end{subfigure}
\begin{subfigure}[b]{.38\textwidth}
\centering
\includegraphics[width=1.5 in,height=1.5 in, trim=0 4 0 0, clip=true]{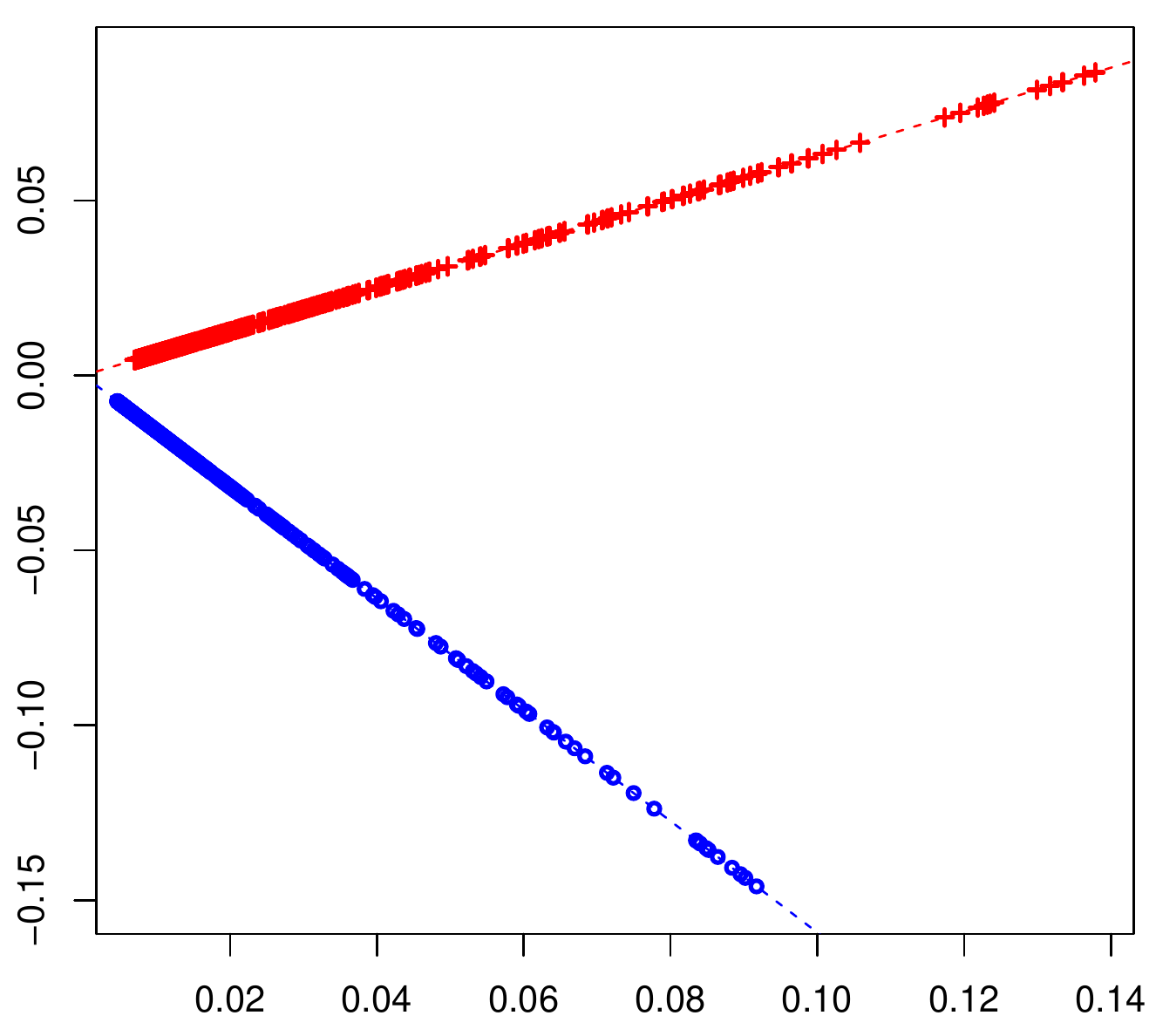}
\caption{Simulated data from the DCBM model.}
\end{subfigure}
\caption{In (a), we plot the rows of $\widehat{\Xi}$ (left) and $\widehat{R}$ (right) for the Polblogs network. In (b), we plot the rows of $\Xi$ using a network simulated from models \eqref{model0a}-\eqref{model0b} ($n = 1222$, $(a, b, c) = (1, 0.1 , 1)$ and $\theta^{-1}_1, \ldots,\theta^{-1}_n$ are iid from $\mathrm{Uniform}([1,20])$), which clearly shows an ideal simplicial cone.}
\label{fig:twocommunities}
\end{figure}

For the polblog data,  since no entry of $\widehat{R}$ exceeds $\log(n)$ in magnitude, $\widehat{R}^* = \widehat{R}$. 
Also, since $K = 2$, the matrix $\widehat{R}$ only has one column and is a vector in $\mathbb{R}$.   
In the left panel of Figure \ref{fig:twocommunities},  we plot the $n$ points $(\hat{\xi}_1(i), \hat{\xi}_2(i))$ for $i = 1, 2, \ldots, n$.  In the middle panel, we plot the $n$ points $(i, \widehat{R}(i))$ for $i = 1, 2, \ldots, n$. 
In both panels, we mark point $i$ with a blue circle if node $i$ is a democrat, and a red cross if it is a republican. 
We observe the following.   
\begin{itemize} 
\itemsep0em
\item {\it (A visible simplicial cone in $\mathbb{R}^2$)}. In the left panel, we have two rays that pass through the origin of $\mathbb{R}^2$, and the region bounded by the two rays is a (simplicial) cone in $\mathbb{R}^2$.   All points (blue and red) fall in the cone region. Most of the blue circles (democrats) fall very close to one ray, and most of the red crosses 
(republicans) fall very close to the other ray.  A small fraction of the points fall in the interior of the cone.  
We also observe that, approximately, the distance between a point and the apex (of the cone) is proportional to the 
degree of the node corresponding to the point (so the points corresponding to small-degree nodes aggregate around the apex). 
\item {\it (A simplex in $\mathbb{R}$)}.  As illustrated in the middle panel, most blue circles (democrat  entries of $\widehat{R}$) concentrate at $2.5$ and most red crosses (republican entries of $\widehat{R}$)  concentrate at $-0.6$.  
This says that if we mark all entries of $\widehat{R}$ on the real line, 
then there is a line segment in $\mathbb{R}$ (which is a simplex in $\mathbb{R}$) with two end points of $-0.6$ and $2.5$, respectively, such that 
most blue circles concentrate at the end point at $2.5$,  most red crosses concentrate at the end point at $-0.6$,  and a small fraction of points fall into the interior of the line segment. 
\end{itemize} 
Now, if we use OSC where we cluster the $n$ rows of $\widehat{\Xi}$ into two groups by directly 
applying the $k$-means,  then we tend to cluster nodes with small degrees to one group, and other nodes to the other. Since a low-degree node can be either democrat or republican, the error rate of OSC for clustering is pretty high (i.e., $437 / 1222$).   However, 
if we use SCORE, where we similarly cluster the $n$ rows of $\widehat{R}$ to two groups with 
the $k$-means, then the degrees no longer have an important role in determining the cluster labels,  and we have a much lower clustering error rate (i.e., $58/1222$).

\cite{SCORE} further explained the simplicial cone and simplex with a theoretical model for the polblog network.  
Note that the network has severe degree heterogeneity (see Table \ref{tab:degree}) 
and that there is no obvious way to model the degrees with just a few parameters.  
For these reasons, for each $1 \leq i \leq n$, we introduce a degree heterogeneity parameter $\theta_i > 0$ 
for node $i$.  Denote an $n \times n$ matrix $\Omega$ by 
\begin{equation} \label{model0a} 
\Omega(i,j) = 
\theta_i \theta_j  \times 
\left\{ 
\begin{array}{ll} 
  a, &\qquad \mbox{if $i, j$ are both democrats},   \\ 
  c, &\qquad \mbox{if $i, j$ are both republicans},  \\
  b, &\qquad \mbox{otherwise}, \\ 
\end{array}
\right. 
\end{equation}
where $a, b, c > 0$ are (unknown) scalars. 
Recall that the adjacency matrix $A$ in (\ref{DefineA}) is symmetric, with all entries on the main diagonal 
being $0$.  We assume the upper 
triangle entries of $A$ are independent Bernoulli random variables satisfying 
\begin{equation} \label{model0b} 
\mathbb{P}(A(i,j) = 1) = \Omega(i,j), \qquad 1 \leq i < j \leq n.  
\end{equation} 
Let $\Xi = [\xi_1, \xi_2, \ldots, \xi_K]$ where similarly $\xi_k$ is the $k$-th eigenvector of $\Omega$, and 
let $R$ be the matrix satisfying $R(i,k) = \xi_{k+1}(i) / \xi_1(i)$, $1 \leq k \leq K-1$,  $1 \leq i \leq n$. By Perron's theorem \citep{HornJohnson}, all entries of $\xi_1$ have the same sign, so without loss of generality, 
we assume all entries of $\xi_1$ are positive. 
Under some mild regularity conditions, we expect to see that 
\begin{equation} \label{eigenApprox} 
\widehat{\Xi} \approx \Xi, \qquad \mbox{and} \qquad \widehat{R} \approx R. 
\end{equation} 
We now derive an explicit formula for $\Xi$ and $R$. 
Let ${\cal C}_1$ be the set of all indices for democrat nodes  and ${\cal C}_2$  be 
the set of all indices for republican nodes. Introduce 
$d_1 = [(\sum_{i \in {\cal C}_1} \theta_i^2) / (\sum_{i = 1}^n \theta_i^2)]^{1/2}$ and $d_2 = [(\sum_{i \in {\cal C}_2} \theta_i^2) / (\sum_{i = 1}^n \theta_i^2)]^{1/2}$. Note that $0 \leq d_1, d_2 \leq 1$ and $d_1^2 + d_2^2 = 1$. 
Let 
\begin{equation} \label{eigvec1a} 
\left[
\begin{array}{cc} 
a d_1^2 &  b d_1 d_2 \\
b d_1 d_2 & c d_2^2 \\ 
\end{array} 
\right] =  Q \Lambda Q'  \; \mbox{be the spectral decomposition, where $\Lambda$ is diagonal and $Q$ is orthogonal,}
\end{equation}
 and we write
 \[
 Q = \left[
\begin{array}{cc} 
u_1  & u_2  \\
v_1  & v_2 \\ 
\end{array}
\right].  
\]  
Write $\Xi = [x_1, x_2, \ldots, x_n]'$ so that $x_i$ is row $i$ of $\Xi$.  In the current case, $K = 2$, so $R$ only has one column.  By basic algebra, we have 
\[
\mbox{The $i$-th row of $\Xi$}   =   \left\{
\begin{array}{ll}
\theta_i  (d_1 \|\theta\|)^{-1}   (u_1, u_2),  &\qquad \mbox{if node $i$ is a democrat},     \\
\theta_i  (d_2 \|\theta\|)^{-1}   (v_1, v_2),    &\qquad \mbox{if node $i$ is a republican}.  \\
\end{array}
\right. 
\]
and  
\[
R(i) = \left\{
\begin{array}{ll}
u_2/u_1,  &\qquad \mbox{if node $i$ is a democrat},     \\
v_2/v_1,    &\qquad \mbox{if node $i$ is a republican}.  \\
\end{array}
\right.
\] 
Combining these with (\ref{eigenApprox}),  in the two-dimensional plane with $x$ and $y$ being the two axes,  
we have the following. 
\begin{itemize}  
\item In the right half of the plane, there is a simplicial cone bounded by two rays $y = (u_2 / u_1) x$ and $y = (v_2 / v_1) x$.  For $1 \leq i \leq n$, if node $i$ is democrat, then row $i$ of $\widehat{\Xi}$ falls close to one of the two rays above, 
and its distance from the apex of the cone is (approximately) proportional to 
the degree heterogeneity parameter $\theta_i$. Similarly, if node $i$ is republican, then row $i$ falls close to the other ray,  and its distance from the apex of the cone is proportional to $\theta_i$ approximately.
If we cluster the $n$ rows of $\widehat{\Xi}$ into two groups with $k$-means, we tend to cluster low-degree nodes into one group. Since a low-degree node may be either democrat or republican, we end up with many clustering errors. 
\item The SCORE normalization transforms the simplicial cone above to a line segment in $\mathbb{R}$ (which can be viewed as a simplex in $\mathbb{R}$),  where the 
two end points are the slopes of the two rays above, respectively.  The transform largely removes the ancillary effect of the parameters $\theta_i$, and as a result,  the democrat rows of $\widehat{R}$ concentrate on one endpoint, and 
the republican rows concentrate on the other.  If we similarly cluster the $n$ rows of $\widehat{R}$ into two groups by $k$-means, we expect much lower error rates than that above. 
\end{itemize} 
This explains what we observe for the polblog network above.   Also see \cite[Section 1.1]{SCORE} for the explicit formula for $u_1, u_2, v_1, v_2$.   Model (\ref{model0a})-(\ref{model0b}) is for a simple case with only two communities, 
but our findings for SCORE and the simplex structure are valid for heterogeneous networks with multiple communities. 
See details below.

\subsection{SCORE and the simplex structure in more general settings}  
In the simple model (\ref{model0a})-(\ref{model0b}), we do not allow mixed-membership (e.g., a web blog is partially democrat and partially republican).   For networks with mixed-memberships, we similarly have a simplicial cone in $\mathbb{R}^K$ associated with the rows of $\Xi$. The SCORE normalization removes the effect of degree heterogeneity and converts the cone to a simplex in $\mathbb{R}^{K-1}$. However, in this case, it may happen that a row of $\widehat{R}$ falls deeply into the interior of the simplex; this will be elaborated in Section \ref{subsec:idealsimplexDCMM}.  Also, the SCORE normalization provides a simple and direct link between the simplex and the network memberships. Utilizing this link gives rise to an approach called Mixed-SCORE, which can be viewed as an adaption of SCORE for networks with mixed-memberships (such networks are harder to analyze than the networks without mixed-memberships).  

SCORE and Mixed-SCORE are for heterogeneous network data. 
Interestingly, for heterogeneous text data,  SCORE 
continues to be useful. With a latent variable model for text documents, we discover two similar simplicial cones in the spectral domain,  each of which can be transformed to a simplex using a step that is similar to the SCORE normalization. This 
gives rise to Topic-SCORE as a new approach to topic learning.    

Additional to Mixed-SCORE and Topic-SCORE, we also have many other variants of 
SCORE, including SCORE+, Hier-SCORE, Tensor-SCORE, and Dynamic Mixed-SCORE.  
With an asymptotic framework, we review precise theory for these methods.  
We show that SCORE attains an exponential rate and a sharp phase transition in community detection, and that Mixed-SCORE and Topic SCORE are rate-optimal in broad settings.

\subsection{Two data sets on the publications of statisticians} 
\label{subsec:twodata}  
SCORE is useful in many applications. For example, SCORE was successfully applied to a New York Taxi data set \citep{duan2019state}, 
an international trade network \citep{MSCORE}, and all $8$ data sets in Table \ref{tab:degree}.  SCORE is especially useful for analyzing  two recent data sets (Phase I and Phase II) on the publications 
of statisticians. 
The Phase I data set was reported in \cite{SCC-JiJin}. 
It consists of the bibtex (e.g., title, author, abstract, reference) and citation data of 3248 published papers (of a total of 3607 authors)  in $4$ journals (Annals of Statistics, 
Biometrika, Journal of the American Statistical Association, and Journal of the Royal 
Statistical Society, Series B) from 2003 to 2012. 
The Phase II data was reported in \cite{PhaseII-JBES,PhaseII-AOAS}. It 
contains the Phase I data set as a subset and consists 
 of the citation and bibtex data of 
83,331 published papers in $36$ journals in statistics and related fields 
from 1975 to 2015.  See the table below.    
The data sets provide a fertile ground for research in bibliometrics and statistics. For example, using the data sets,  we can construct 
many different types of networks  (e.g., co-authorship network, 
citation network,  author-paper bipartite network).  Also, the data sets contain the abstract and references for each paper, which can be used as text documents for text mining.  The study on the data set provides   valuable insight on understanding the 
research interests and habits of statisticians and may help the administrators (e.g., 
universities, funding agencies) in decision making.

\begin{center}
\scalebox{0.9}{ 
\begin{tabular}{lrrrrr}
\hline
 &  \#journals &    time span & \#authors & \#papers  \\
\hline
Phase I   & $4$   & $2003$-$2012$  &  $3,607$  & $3, 248$   \\ 
Phase II  &  $36$    &  $1975$-$2015$   &  $47,311$  &  $83,331$  \\
\hline
\end{tabular}
}
\label{tab:twodata} 
\end{center}

\section{Community detection by SCORE and several variants of SCORE}  \label{sec:SCORE}   
Given a network with $K$ perceivable communities, the goal of 
community detection is to assign each node to (exactly) one of the $K$ communities 
(i.e., node clustering).  For community detection, it is conventional to assume 
that (at least in theory) none of the nodes have mixed memberships, and the {\it Degree-Corrected 
Block Model (DCBM)} below  is an appropriate model. 
For broader settings where a non-negligible fraction of nodes have mixed-memberships, 
it is preferable to use the broader {\it Degree-Corrected 
Mixed-Membership (DCMM)} model (see below).  In such a case, 
node clustering is no longer the proper goal, so 
we consider the problem of membership estimation 
(which includes community detection as a special case). 
In this section, we discuss the problem of community detection focusing on the DCBM model.  
The discussion for membership estimation is deferred to Section \ref{sec:MME}, 
where we focus on the DCMM model.  We start by introducing these two models. 
\subsection{The DCMM model, with the DCBM as a special case} 
\label{subsec:DCMMmodel} 
The DCMM model \citep{JiZhuMM, MSCORE, JKL2021} is a recent network model, 
which aims to accommodate two noteworthy features in 
real networks: severe degree heterogeneity and mixed memberships.  
Consider an undirected and connected network with $n$ nodes and let $A$ be the adjacency matrix as in (\ref{DefineA}).  
Suppose that the network has $K$ perceivable communities 
${\cal C}_1, {\cal C}_2, \ldots, {\cal C}_K$ (communities in a network are tightly woven groups of nodes that have 
more edges within than between). For each $1 \leq i \leq n$, we model the membership vector of node $i$ by 
a weight vector $\pi_i \in \mathbb{R}^K$ (we call $\pi \in \mathbb{R}^K$ a weight vector if all entries are non-negative with a unit sum) where 
\begin{equation} \label{DCMM1} 
\pi_i(k)  = \mbox{weight node $i$ puts in community $k$}, \qquad 1 \leq k \leq K.  
\end{equation}

\begin{definition}
We call node $i$ a pure node if $\pi_i$ is degenerate (i.e., one entry is $1$, other entires are $0$) and a mixed node otherwise. 
When node $i$ is pure, there is a unique $k$, $1 \leq k \leq K$,  such that $\pi_i(k) = 1$ and  
$\pi_i(m) = 0$ for $m\neq k$.  In this case, we call node $i$  a pure node of community $k$.   
\end{definition}
For a non-negative and symmetric matrix $P \in \mathbb{R}^{K, K}$ that models the community structure, assume the upper triangle entries of $A$ are independent Bernoulli random variables such that 
\begin{equation} \label{DCMM2} 
\mathbb{P}(A(i,j) = 1) = \theta_i \theta_j \cdot \pi_i' P \pi_j. 
\end{equation} 
For identifiability  we assume 
\begin{equation} \label{DCMM3} 
\begin{array}{l}
\mbox{$P$  is non-singular and has unit diagonal entries, and each}\\
\mbox{ of the $K$ communities has at least one pure node}.   
\end{array}
\end{equation} 
Here, we say model (\ref{DCMM2})  is identifiable if $P$ and $\{(\theta_i, \pi_i)\}_{i = 1}^n$ are uniquely determined by $\mathbb{E}[A]$.  Without (\ref{DCMM3}),  model (\ref{DCMM2}) may not be identifiable.   
The identifiability was studied in several recent papers (e.g., \cite{MSCORE,JKL2021,jin2020estimating}). 
There are many ways to make model (\ref{DCMM2}) identifiable with a mild condition, and among them, (\ref{DCMM3}) is a convenient choice.  We call (\ref{DCMM1})-(\ref{DCMM3}) the DCMM model.   
Introduce $\theta = (\theta_1, \theta_2, \ldots, \theta_n)'$,  $\Theta = \diag(\theta_1, \theta_2, \ldots, \theta_n)$,  
and $\Pi = [\pi_1, \pi_2, \ldots, \pi_n]'$. 
We can rewrite (\ref{DCMM2}) as
\begin{equation} \label{DCMMmatrix} 
A = \Omega - \mathrm{diag}(\Omega) +W = ``\mbox{main signal}" + ``\mbox{secondary signal}" + ``\mbox{noise}",
\end{equation}
where 
\[
\Omega = \Theta \Pi P \Pi' \Theta\quad \mbox{and} \quad W = A - \mathbb{E}[A].   \;\;\; \footnote{For a vector $x \in \mathbb{R}^n$, $\diag(x)$ stands for the $n \times n$ diagonal matrix $X$ with $X(i,i) = x_i$, $1 \leq i \leq n$. For an $n \times n$ matrix $X$, $\diag(X)$ stands for the $n \times n$ diagonal matrix where $(\diag(X))(i,i) = X(i,i)$, $1 \leq i \leq n$.}   
\]
 
\begin{figure}[tb!]
\begin{center}
\includegraphics[height = .75 in]{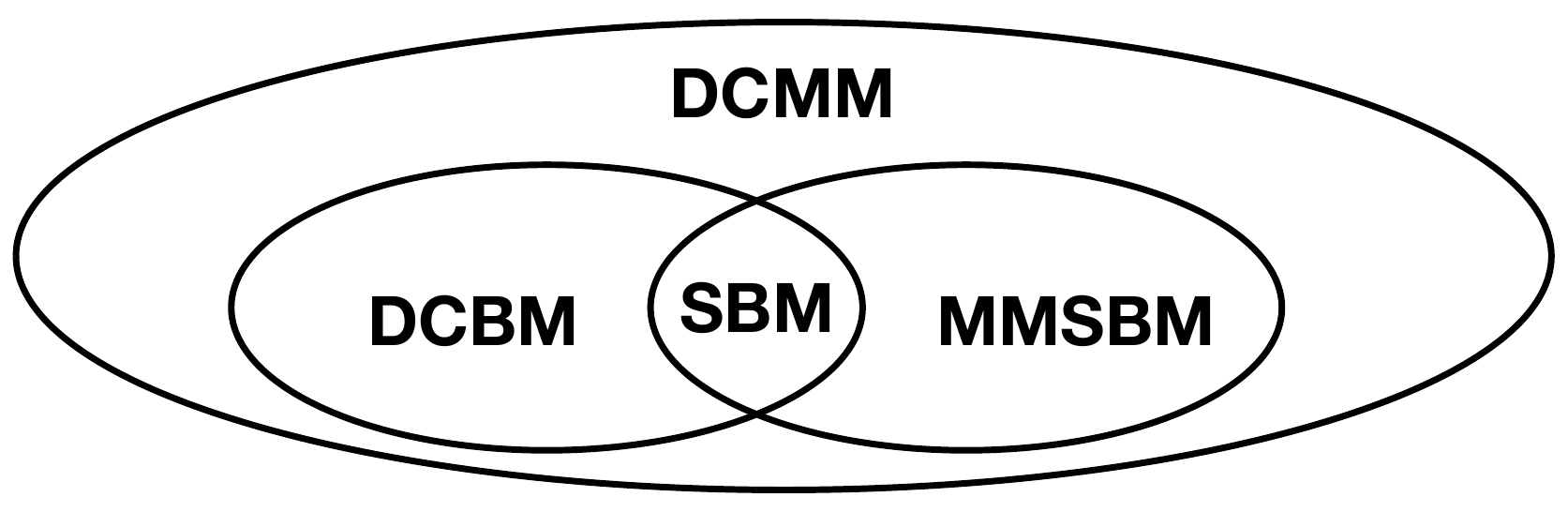}
\caption{Comparison of four models (DCMM, DCBM, MMSBM, and SBM) by the Venn diagram.} 
\label{fig:model}
\end{center} 
\end{figure}

Note that all off-diagonal entries of $W$ are zero-mean centered Bernoulli  variables, so $W$ is a generalized Wigner matrix \citep{genWigner}.  For most range of interests,  $\diag(\Omega)$ was found to only have a negligible effect 
asymptotically (e.g., \cite{SCORE, JKL2021}), so we call $\Omega$ and $\diag(\Omega)$ the main signal and secondary signal, respectively.  DCMM was first introduced by \cite{JiZhuMM} with a different name, where each $\pi_i$ is  normalized to have a unit $\ell^2$-norm, so it is unclear how to interpret them. \cite{MSCORE, JKL2021} proposed to normalize each $\pi_i$ to have a unit $\ell^1$-norm, and interpret them as weight vectors in (\ref{DCMM1}). By basic algebra, the two versions of DCMM models are equivalent: we can convert one to the other by re-scaling $(\theta_i, \pi_i)$ for each $1 \leq i \leq n$, while keeping the matrix $P$ unchanged.

DCMM includes several well-known models as special cases. For example, when all nodes are pure (i.e., mixed-membership is not allowed), all weight vectors $\pi_i$ are degenerate, and DCMM reduces to the {\it Degree-Corrected Block Model (DCBM)} \citep{DCBM}. DCBM does not allow mixed memberships, so  for each $1 \leq i \leq n$, there is a unique community $1 \leq k \leq K$ such that node $i$ belongs to community ${\cal C}_k$.  
If we do not allow degree heterogeneity by  requiring $\theta_1 = \theta_2 = \ldots =  \theta_n$ (but the matrix $P$ is not required to have unit diagonal entries),  then DCMM reduces to the Mixed-Membership Stochastic Block Model (MMSBM) \citep{airoldi2009mixed}. If we further require all $\pi_i$'s are degenerate, then MMSBM reduces to the Stochastic Block Model (SBM) \citep{holland1983stochastic}.   See Figure \ref{fig:model} for a comparison of all four models.  
Also, see \cite{sengupta2018block, noroozi2021estimation, chen, yuan2021community} for different but closely related 
network modeling strategies.

Similar to DCMM, DCBM has many parameters, so there is also an identifiability issue. 
But since it is a special DCMM, DCBM is identifiable if we assume that $P$  is non-singular and  has unit diagonals  
(in this case,  the second part of (\ref{DCMM3}) holds automatically; also, sometimes,  we can drop the non-singularity requirement on $P$).  
The MMSBM and SBM have relatively few parameters, and they are identifiable without the constraint in (\ref{DCMM3}).

\subsection{The ideal simplicial cone and simplex for DCBM} \label{subsec:idealsimplexDCBM}   
In Section \ref{subsec:SCOREbasics}, we discuss a special DCBM and show that there is 
a simplicial cone and a simplex in the spectral domain (the model does not 
satisfy (\ref{DCMM3}), but we can re-parameterize it to satisfy (\ref{DCMM3})).  
We now discuss the simplicial cone and the simplex for general DCBM. The discussion for the more general DCMM model is 
deferred to Section \ref{subsec:idealsimplexDCMM}.   
Let $\Omega = \Theta \Pi P \Pi' \Theta$ be the main signal matrix as in (\ref{DCMMmatrix}). Similar to Section \ref{sec:intro},  for $1 \leq k \leq n$, let $\lambda_k$ be the $k$-th largest (in magnitude) eigenvalue of 
$\Omega$ and let $\xi_k$ be the corresponding eigenvector. By Perron's theorem \citep{HornJohnson}, 
we can similarly assume all entries of $\xi_1$ are strictly positive.   
Introduce $\Xi \in \mathbb{R}^{n, K}$ and $R \in \mathbb{R}^{n, K-1}$ by $\Xi = [\xi_1, \xi_2, \ldots, \xi_K]$ and 
$R(i,k) = \xi_{k+1}(i) / \xi_1(i)$, $1 \leq k \leq K-1$, $1 \leq i \leq n$. 
The two matrices are the non-stochastic counterparts of $\widehat{\Xi}$ and $\widehat{R}$ respectively.
Write 
\[
\Xi = [x_1, x_2, \ldots, x_n]',  \qquad \mbox{and} \qquad   
R = [r_1, r_2,\ldots, r_n]',  
\]
so $x_i'$ and $r'_i$ are the $i$-th row of $\Xi$ and $R$, respectively, $1 \leq i \leq n$. By basic algebra, there is a matrix  $B \in \mathbb{R}^{K,K}$ such that $\Xi = \Theta \Pi B$. 
As in Section \ref{subsec:SCOREbasics}, the community labels (quantities of interest) are contained in the matrix $\Pi$.  The degree heterogeneity parameters $\theta_i$ present as a nuisance, but  can be removed by the 
SCORE normalization.  In detail, write $B = [b_1, b_2, \ldots, b_K]$ and introduce a matrix $V = [v_1, v_2, \ldots, v_K] \in \mathbb{R}^{K-1, K}$ by 
\[
v_k(\ell) = b_{\ell+1}(k) / b_1(k),  \qquad 1 \leq \ell \leq K-1, 1 \leq k \leq K. 
\] 
By basic algebra, we have 
$R = \Pi V'$.  
\begin{definition} 
Let $C(b_1, \ldots, b_K)  = \{\beta_1 b_1 +  \ldots +\beta_K b_K: \beta_k \geq 0\}$ be the simplicial cone in $\mathbb{R}^K$ spanned by the vectors $b_1, b_2, \ldots, b_K$, and 
let $S(v_1, \ldots, v_K)=\{\beta_1v_1+\ldots+\beta_Kv_K: \beta_k\geq 0, \sum_{k=1}^K \beta_k=1\}$ be the simplex in $\mathbb{R}^{K-1}$ with 
$v_1, v_2, \ldots, v_K$ being the vertices. 
\end{definition}  
Recall that under DCBM, each row of $\Pi$ is a degenerate weight vector, and 
each node is a pure node. Theorem \ref{thm:idealsimplexDCBM} is adapted from \cite{SCORE}.
\begin{thm} \label{thm:idealsimplexDCBM} 
{\bf (Ideal simplicial cone and ideal simplex for DCBM}).  Under the DCBM, each node is a pure node, and the minimum 
distance between the $K$ vertices of $S(v_1, v_2, \ldots, v_K)$ is no less than $\sqrt{2}$. 
If node $i$ is a pure node of community $k$, then row $i$ of $\Xi$ falls on the $k$-th ray of $C(b_1, b_2, \ldots, b_K)$ and row $i$ of $R$ falls on the $k$-th vertex of $S(v_1, v_2, \ldots, v_K)$.  The $n$ rows of $R$ take exactly $K$ distinct values. If we partition all the rows to $K$ groups according to the $K$ distinct values, then we  recover the community labels of all $n$ nodes exactly. 
\end{thm} 
In such an oracle case, directly applying $k$-means to the $n$ rows of $\Xi$ 
may result in many errors, but directly applying $k$-means to the $n$ rows of 
$R$ fully recovers all true community labels. 
It is therefore highly desirable to convert the simplicial cone to the simplex by the SCORE normalization.   
The discussion here is for the oracle case where $\Omega$ is accessible. In the real case, $\Omega$ is not accessible, 
but the idea is easily extendable if we replace $R$ by $\widehat{R}^*$ in Section \ref{subsec:SCOREbasics}: this gives rise to the SCORE algorithm we introduced in Section \ref{subsec:SCOREbasics}. 
The simplicial cone and simplex here are reminiscent of the simplicial cone in \citep{donoho2003does} for non-negative matrix factorization, but are for  very different settings. 

\subsection{The Laplacian normalization and comparison of pre-PCA and post-PCA approaches}  
\label{subsec:Laplacian} 
SCORE is a so-called {\it post-PCA normalization} approach, 
for the normalization is applied to the leading eigenvectors or principal components of the data matrix (e.g., \cite{Topic}).  
It is quite different from the Laplacian approach in  \cite{rohe2011}, which can be viewed as a pre-PCA normalization approach for  we apply the Laplacian normalization before we compute the principal components.  Let $d_i$ be the degree of node $i$, $1 \leq i \leq n$. Let $D = \diag(d_1, d_2, \ldots, d_n)$ and let $\bar{d}$ be the average of $d_1, d_2,\ldots, d_n$. For a ridge parameter $\delta > 0$, the Laplacian approach runs as follows (the approach in \cite{rohe2011} corresponds to $\delta = 0$). 
\begin{itemize} 
\itemsep0em
\item Obtain the Laplacian normalization matrix $\widetilde{A} = (D + \delta \bar{d} I_n)^{-1/2} A (D + \delta \bar{d} I_n)^{-1/2}$.  
\item Obtain the first $K$ eigenvectors of $\widetilde{A}$ and arrange them in an $n \times K$ matrix. 
\item Cluster the $n$ nodes into $K$ groups by applying the $k$-means to the $n \times K$ matrix above.  
\end{itemize} 
Just like SCORE, the Laplacian normalization (step 1 above) also aims to alleviate 
the effect of degree heterogeneity, but there is a problem. 
Recall that the adjacency matrix 
$A$ can be decomposed into three components: $A = \Omega - \diag(\Omega) + W$, where $\Omega$ is the main signal component and $W$ is the noise component.  
For $1 \leq i \neq j \leq n$,  the mean and standard deviation of $W(i,j)$ are $0$ and $[\Omega(i,j) (1 - \Omega(i,j))]^{1/2}$, respectively.  For most networks and most $(i, j)$,  $0< \Omega(i,j) \ll 1$,    so 
the standard deviation of $W(i,j)$ is at the order of $[\Omega(i,j)]^{1/2}$ which is much larger than $\Omega(i,j)$. 
Therefore,  a proper normalization for the ``signal'' part may not be suitable for the ``noise'' part, and vice versa. As a result, to alleviate the effect of degree heterogeneity,   we need quite different normalizations for $\Omega$ and $W$.  
This is challenging, as both $\Omega$ and $W$ are unknown.  
The Laplacian approach uses the same normalization for the 
main signal matrix and the noise matrix. This is a convenient choice,  but 
it does not overcome the challenge.  As a result, while the approach is shown to be consistent under SBM (a special case of DCBM where all $\theta_i$'s are equal; see Figure \ref{fig:model}), it is not consistent under the more general DCBM.

\cite{gulikers2017spectral} proposed an alternative pre-PCA normalization approach (which we may call GLM)  where 
they normalize the adjacency matrix $A$ to $\widetilde{A} = (D + \delta I_n)^{-1} A (D + \delta I_n)^{-1}$.  
To see the rationale, consider a DCBM with $K = 1$. In this case,   $\Omega = \theta \theta'$ and $\mathbb{E}[D] \propto \Theta$ approximately, so 
 approximately,  $D^{-1} \Omega D^{-1} \propto {\bf 1}_n {\bf 1}_n'$. Compared to the Laplacian approach, GLM is more reasonable 
if we only need to normalize $\Omega$. Unfortunately, as we need to normalize 
both $\Omega$ and $W$,  GLM faces the same challenge as the Laplacian approach.  

SCORE overcomes the challenge by a very different 
strategy: it first reduces the noise using a PCA step and then normalizes the principal components  of $A$ (which are natural estimates of the principal components of $\Omega$). For this reason,  
SCORE and other post-PCA normalization approaches are expected to be more efficient in alleviating the effect of degree heterogeneity than the pre-PCA normalization approaches.

Naturally, pre-PCA and post-PCA can be combined.  For example,  the RSC approach by \cite{qin2013regularized}  is the direct combination of SCORE and the Laplacian normalization above with $\delta = 0$ (RSC is the same as an earlier approach proposed in \cite[Section 1.6]{SCORE}. Note that while \cite{SCORE} is published in 2015, the arXiv version 
appeared much earlier in November 25, 2012). 

In Table \ref{tab:error}, we compare SCORE with the Laplacian approach (Lap0, Lap1, for $\delta = 0$ and $.05$ respectively), the GLM approach (GLM0 and GLM1, for $\delta = 0$ and $.05$ respectively), and the RSC approach.  The GLM approach and the Laplacian approach are seen to underperform both SCORE and RSC, especially for the Caltech, polblogs and Simmons data sets (which are the more difficult ones among the $8$ data sets).   Also, RSC  does not show a significant improvement over SCORE (in fact, for $3$ of the data sets, polblogs, dolphins and polbooks, RSC slightly underperforms SCORE). A possible reason is that,  SCORE is already effective in alleviating the effect of degree heterogeneity,  
so adding the Laplacian normalization does not provide any substantial improvement. 
Of course, in some other settings,  it could happen that the SCORE normalization alone is inadequate, and it is desirable to use both. See Section \ref{sec:TM}, where we review topic-SCORE  \citep{Topic}, a topic modeling approach to analyzing text documents. In this approach, both pre-PCA and post-PCA normalizations play an important role.

\subsection{Other SCORE-type post-PCA normalization approaches}   
\label{subsec:SCORE-type} 
In \cite[Supplement A]{SCORE}, a family of SCORE-type post-PCA normalization approaches were proposed.  
We call a mapping $M$ from $\mathbb{R}^K$ to $[0, \infty)$ a scale-invariant mapping if for any $x \in \mathbb{R}^K$ and $c > 0$,   $M(cx)=c M(x)$.  Given any scale-invariant mapping $M$, we define a normalization by dividing $x_i$, the $i$th row of $\Xi$, by $M(x_i)$. Writing $r_i=x_i/M(x_i)\in\mathbb{R}^K$, we can similarly construct $\widetilde{R}=[r_1,r_2,\ldots,r_n]'$. This includes the SCORE normalization as a special case,  where $M(x)=x(1)$ so we divide each row of $\Xi$ by its first entry.  In this case, the first column of $\widetilde{R}$ is the vector of all ones and so is not informative;  removing it gives rise to the matrix $R$ in Section~\ref{subsec:idealsimplexDCBM}. Other choices of $M$ include $M(x)=\|x\|_q$, for some $q>0$. This is called the SCORE$_q$ normalization, which normalizes each row of $\Xi$ by the $\ell^q$-norm of that row. Later in the literature, a similar idea is used in \cite{lei2015consistency}, which corresponds to the case of $q=2$, and a similar idea is used in \cite{gao2016community} which corresponds to the case of $q = 1$ (they constructed a rank-$K$ approximation $\widehat{A}$ to $A$ and normalized each row of $\widehat{A}$ by the $\ell^1$-norm of this row; since $\widehat{A}$ is already a low-rank matrix, this is essentially a post-PCA normalization). The case of $q=2$ is also similar to that in \cite{ng2002spectral}. Note that the setting in \cite{ng2002spectral} is different from ours, and it does not provide theoretical justifications (e.g., the simplex structure) 
as we do.  

Compared to SCORE$_q$,  the SCORE normalization has several advantages. First,  fixing $1 \leq i \leq n$, we compare the Signal to Noise Ratio (SNR) in $\hat{\xi}_1(i), \hat{\xi}_2(i), 
\ldots, \hat{\xi}_K(i)$. It is known that the SNR in $\hat{\xi}_k(i)$ decreases with $k$. As a result,  the $\ell^q$-norm 
of the vector $(\hat{\xi}_1(i), \hat{\xi}_2(i), \ldots, \hat{\xi}_K(i))$ is noisier than 
$\hat{\xi}_1(i)$, so the SCORE$_q$ normalization may introduce more noise than the SCORE normalization.  
Second, in practice, the number of communities $K$ is unknown, and  we need to run SCORE with $K = 5, 6, \ldots, 15$ (say) and see which $K$ gives the most reasonable results (e.g., \cite{PhaseII-JBES}).  
For either of the two normalization approaches, let $\widehat{R}^{(5)}, \widehat{R}^{(6)}, \ldots, \widehat{R}^{(15)}$ be the corresponding matrix $\widehat{R}$. If we use the SCORE normalization,  then an appealing property is that, 
the $\widehat{R}^{(5)}$ equals  to the submatrix of $\widehat{R}^{(6)}$ excluding the last column,  $\widehat{R}^{(6)}$ equals to the submatrix of $\widehat{R}^{(7)}$ excluding the last column, and so on and so forth. If we use the 
SCORE$_q$ normalization, such a property does not hold. 
Last and most importantly, in Section \ref{sec:MME}, we show that when the network has mixed-memberships, the SCORE normalization converts the ideal simplicial cone to a simplex and provides a simple and direct link between the simplex and memberships. If we use SCORE$_q$, we do not have such a simple and direct link; see Section \ref{subsec:postPCADCMM}.

\subsection{SCORE+, an improvement over SCORE for networks with weak signals}  \label{subsec:SCORE+}   
For most of the data sets in Table \ref{tab:error}, SCORE performs well. 
For two of the data sets (Simmons and Caltech) there, 
SCORE underperforms (e.g., compared to CMM and LSCD, to be introduced in Section \ref{subsec:otherCD}).  \cite{SCORE+} pointed out that this is because 
these two networks have relatively {\it weak signals} while the other $6$ data sets 
have relatively {\it strong signals}.  See Table \ref{tab:error} for example, 
where for Simmons and Caltech, the error rates of all methods are relatively high.  
This is is reminiscent of that in \cite{DJ15}.  
\cite{SCORE+} further proposed SCORE+ as an improvement 
of SCORE especially for networks with weak signals. The algorithm is as follows (we recommend $t = 0.1$ and $\delta =  0.05$ or $\delta = 0.1$): 
\begin{itemize} \itemsep 0em
\item {\it (Pre-PCA normalization with Laplacian).}  Let  $D = \diag(d_1, d_2, \ldots, d_n)$ where $d_i$ is the degree of node $i$. Obtain the regularized graph Laplacian as  $L_{\delta} = (D + \delta \cdot d_{max} \cdot  I_n)^{-1/2} A (D + \delta \cdot d_{max} \cdot  I_n)^{-1/2}$, where $d_{max} = \max_{1 \leq i \leq n} \{d_i\}$ and $\delta>0$ is the ridge regularization parameter. 
\item {\it (PCA, with possibly an additional eigenvector).}  Obtain $(\hat{\lambda}_k^*, \hat{\xi}_k^*)$ for $k = 1, \ldots, K+1$, 
where $\hat{\lambda}_k^*$ is the $k$-th largest (in magnitude) eigenvalue of $L_{\delta}$ and $\hat{\xi}_k^*$ is the corresponding eigenvector. 
For a given threshold $t>0$, we set $M=K+1$ if $(\hat{\lambda}_K^*-\hat{\lambda}_{K+1}^*) /\hat{\lambda}_{K}^* \leq t$ and $M=K$ otherwise. 
\item {\it (Post-PCA normalization).}  Let $\hat{\eta}_k=\hat{\lambda}_k^* \hat{\xi}_k^*$, for $1\leq k\leq M$. Obtain the matrix $\widehat{R}$ by $\widehat{R}(i,k)=\hat{\eta}_{k+1}(i)/\hat{\eta}_1(i)$, $1\leq i\leq n, 1\leq k\leq K-1$. 
\item {\it (Clustering)}. Apply classical $k$-means to  the rows of  $\hat{R}$, assuming $\leq K$ clusters. 
\end{itemize}

As mentioned in Section \ref{subsec:Laplacian}, 
we can always combine a Laplacian pre-PCA step with 
SCORE. This explains step 1.  
We now explain step 2-3.  For simplicity, let us remove the Laplacian regularization in Step 1, so $L_{\delta} = A$, and $(\hat{\lambda}_k^*, \hat{\xi}_k^*) = (\hat{\lambda}_k, \hat{\xi}_k)$, where we recall $\lambda_k$ and $\hat{\lambda}_k$ are the $k$-th largest (in magnitude) eigenvalue of $\Omega$ and $A$, respectively, and $\xi_k$ and $\hat{\xi}_k$ are the corresponding eigenvectors. 
Consider step 3 first.  Recall that in Section \ref{subsec:SCORE-type},  for each $1 \leq i \leq n$, 
the SNR in $\hat{\xi}_k(i)$ decreases with $k$. Therefore, before we apply $k$-means to 
$n$ rows of the matrix $[\hat{\xi}_1, \hat{\xi}_2, \ldots, \hat{\xi}_M]$,  it is preferable to 
down weight $\hat{\xi}_k$ as $k$ increases.  Since the noise level in $\hat{\xi}_k$, measured by the $\ell^2$-norm error, is approximately proportional to $1/\lambda_k$, an appropriate choice for down weighting is  to multiply each 
$\hat{\xi}_k$ by $\hat{\lambda}_k$. This explains step 3. 
 It remains to explain step 2. 
Our numerical study finds that, although step 1 and 3 may offer some improvements over SCORE, the improvements are usually not significant; see our discussion on RSC in Section \ref{subsec:Laplacian}.   
 In order for SCORE+ to have a significant improvement,  it is crucial to use step 2, where we {\it may}  include one more eigenvector before we apply the $k$-means for clustering.  This is motivated by the following observation.  
Since the rank of $\Omega$ is $K$, it has only $K$ nonzero eigenvalues, $\lambda_1, \ldots, \lambda_K$.   
In the strong signal case, $|\lambda_K|$ is much larger than the noise level, measured by the spectral norm of the noise matrix, $\|W\|$. In this case, by spectral analysis, $|\hat{\lambda}_{K}|$ is much larger than $|\hat{\lambda}_{K+1}|$ and $\hat{\xi}_K$ is highly correlated with $\xi_K$,  while 
$\hat{\xi}_{K+1}$ is only weakly correlated with $\xi_K$. 
In the weak signal case, $|\lambda_K|$ may be  close to or even smaller than $\|W\|$.  When this happens,  
$\hat{\lambda}_K$ and $\hat{\lambda}_{K+1}$ are very close to each other, and 
$\hat{\xi}_{K+1}$ may be more correlated with $\xi_K$ than $\hat{\xi}_K$. 
Therefore, if $|\hat{\lambda}_K - \hat{\lambda}_{K+1}| / |\hat{\lambda}_K|$ is 
large, we are confident that we are in the strong signal case, and there is no need to 
include one more eigenvector before we apply the $k$-means clustering. If $|\hat{\lambda}_K - \hat{\lambda}_{K+1}| / |\hat{\lambda}_K|$ is relatively small,  then we may be in the weak signal case, so it is important to include one more eigenvector for the $k$-means clustering.  See \cite{SCORE+} for a detailed explanation. 

\begin{table}[bt!]  
\centering
\hspace*{-8em}
\scalebox{.82}{
\setlength\tabcolsep{3pt}
\begin{tabular}{ c c c|ccc ccc ccc ccc c ccc  c }
Dataset & n & K &SCORE  & SCORE$+$ &   OSC &RSC &Lap1 &Lap0  &GLM1 & GLM0           &$SCORE_1$ &$SCORE_2$ & PPL & DCPPL & Tabu    &  LSCD  &    CMM &  OCCAM  & Refine\\
\hline
Polblogs &1222& 2 & 58  & $\bf{51}$   & 437 & 64&    380   &   590   &   256&578      &      61    &     64 & 374  & 59& 62  &    58 &  62 &  65   & 92\\
Simmons  &1137& 4 &  268 & $\bf{127}$    &  442 & 244 & 334    &  278    &   291&781 &  255    &    253 &   333  & 218& 264  &  134 &  137 &  266  & 249 \\
Caltech   &590& 8 &  183 & \bf{98}   &  224 & 170& 204   &   174     &   319&487  & 177   & 178 & 232 & 153 &172  & 106 & 124  & 189  & 158 \\
Football &110 & 11 & 5& 6&   5 &  5&    6     &      6         &     6       &       6         &     5       &      {\bf 4} & 8 & 6 &11 &  21 &  7 & {\bf 4}  & {\bf 4}\\
Karate &34&  2 & $\bf{0}$ & 1&   $\bf{0}$&  $\bf{0}$& {\bf 0}  &  1  &  11 & 11 &  0  &   {\bf 0}  & {\bf 0} & {\bf 0} &1 & 1 & {\bf  0} & {\bf  0} & 1\\
Dolphins  &62& 2 &   $\bf{0}$ &  2     &  12 & 1  & {\bf 0}   & {\bf 0}   &   20 & 20  &    1   & 1 &{\bf 0} & {\bf 0} & {\bf 0}  & 2 & 1  &  1 & {\bf 0}\\
Polbooks  &92 & 2 & $\bf{1}$ & 2    &  4  & 3 &   2     &   2     &       44   &44      &        3       &      3 & 2 & 2 &2 & 3 & {\bf 1} &  3 & 2\\
UKfaculty &79 & 3 &  2 & 2   & 5  & $\bf{0}$ & 1       &       1     & 3&31     & 5  & 6 & 3  & 2&2  & 1 & 7  &  5 & 2\\
\hline
\end{tabular}
}
\caption{Comparison of clustering errors of 16 community detection approaches on the $8$ datasets in Table~\ref{tab:degree}.   Lap0 and Lap1: Laplacian approach  with $\delta = 0$ and $\delta = .05$;  
GLM0 and GLM1: GLM approach with $\delta = 0$ and $\delta = .05$; see Section \ref{subsec:Laplacian}.   SCORE$_1$ and SCORE$_2$: SCORE$_q$ with $q = 1$ and $q = 2$; see Section \ref{subsec:SCORE-type}.  DCPPL, PPL, Tabu, LSCD, CMM: see Section \ref{subsec:otherCD}. 
OCCAM: see Section \ref{subsec:otherME}. Refine: the approach by running a refinement of SCORE; see Section~\ref{subsec:SCORErate}. Boldface: lowest error among all $17$ methods.} 
\label{tab:error} 
\end{table}

In Table~\ref{tab:error}, we compare SCORE+ with SCORE for $8$ data sets.  
For the two data sets with weak signals (Simmons and Caltech),  SCORE+ improves SCORE significantly. 
If we remove step 2 in SCORE+, there is still an improvement, but it is not as significant.  
For the other $6$ data sets, SCORE+ and SCORE have similar results. This is 
as expected, as the signals in these data sets are relatively strong, so there is not much room 
for improvement.

\subsection{SCORE*, an improvement of SCORE for assortative networks}  
\label{subsec:SCORE*} 
\cite{PhaseII-JBES} analyzed many (e.g., co-authorship, citee) networks of statisticians constructed by the Phase II data set in Table \ref{tab:twodata}.  
When they were analyzing a co-authorship network for statisticians,   
 they discovered that if they directly apply SCORE for community detection, 
then some of the resultant communities are dis-assortative (a community is 
assortative if it has more edges within than between among all of its 
sub-communities).  Since a co-authorship community  is expected to be 
assortative, the authors proposed a small modification for SCORE. 
Recall that in SCORE, we apply $k$-means to the  $n$ rows of the matrix $\hat{R}=[\diag(\hat{\xi}_1)]^{-1}[\hat{\xi}_2, \ldots, \hat{\xi}_K]$, 
where $\hat{\xi}_k$ is the $k$-th eigenvector of $A$. 
The modified version runs in the same way, except that we let $\hat{\xi}_k$ be the $k$-th 
eigenvector of $A + c_0I_n$, for a tuning parameter $c_0 > 0$.  
Though any eigenvector of $A$ is also an eigenvector of $A + c_0 I_n$, the subtle difference here lies in the definition of the ``$k$-th eigenvector": $\hat{\xi}_k$ is the $k$-th eigenvector of $A$ if the corresponding eigenvalue $\hat{\lambda}_k$ is the 
$k$-th largest among all eigenvalues of $A$ {\it in magnitude}, and it is the $k$-th eigenvector of $A + c_0 I_n$ if 
$\hat{\lambda}_k + c_0$ is the $k$-th largest among all eigenvalues of $A + c_0 I_n$ {\it in magnitude}. 
The latter punishes negative eigenvalues by ranking them lower. This is motivated by the observation that for assortative networks,  a negative eigenvalue is more likely to be spurious than a positive eigenvalue.  \cite{PhaseII-JBES} found that if we use the modified SCORE with $c_0 = 1$,  then 
the estimated communities are assortative (from a practical viewpoint, such a result is more appealing). 
For network analysis where we encounter a similar issue, we may use the modified SCORE, 
with the same or a different  tuning parameter $c_0$.  
See Section \ref{subsec:statistics} where we discuss how to use SCORE* to build a 
  community tree (a $4$-layer tree with $26$ leaf communities) for a co-authorship network of statisticians.

\subsection{Several recent  community detection approaches}  \label{subsec:otherCD}  
 \cite{bickel2009nonparametric} proposed the community detection method by maximizing a likelihood modularity. 
\cite{zhao2012consistency} and \cite{wang2020fast}  further developed it to a pseudo likelihood approach,  
where the basic idea is to maximize the pseudo likelihood over all 
eligible parameters $(P, \theta, \Pi)$ in the DCBM, assuming the entries 
of $A$ are Poisson random variables.  
Since given $\Pi$, the maximum over all eligible $(P,\theta)$ has an explicit form, 
we can recast the optimization problem as the optimization of a modularity:   
\[
\widehat{\Pi} =\mathrm{argmax}_{\Pi} \sum_{1\leq k,\ell\leq K} O_{k\ell}\log\Bigl(\frac{O_{k\ell}}{O_kO_{\ell}}\Bigr), \qquad\mbox{where}\quad O_{k\ell}=\sum_{(i,j):\pi_i(k)\text{=}1,\pi_j(\ell)\text{=}1}A(i,j), \quad O_k = \sum_{\ell}O_{k\ell}. 
\] 
The above optimization requires an exhaustive search over the community label matrix $\Pi$ and so is NP hard. 
To address the issue, \cite{zhao2012consistency} proposed a greedy algorithm called the Tabu search,  and   
\cite{wang2020fast} proposed an alternating maximization algorithm by decoupling the membership labels of the rows and columns called DCPPL (they also gave a special version PPL, which is for the SBM model).

\cite{chen2015convexified} proposed the LSCD approach, which uses 
semi-definite programming (SDP). Let $d=(d_1,d_2,\ldots,d_n)'$ where $d_i$ is the degree of node $i$, $1 \leq i \leq n$, and let $\lambda > 0$ be a tuning parameter. The LSCD approach first solves an $n\times n$ matrix $\widehat{Y}$ from  
\[
\widehat{Y} =\mathrm{argmax}_{Y\succeq 0}\;\; \langle Y,\; A-\lambda dd'\rangle, \qquad\mbox{subject to}\quad 0\leq Y(i,j)\leq 1,\quad Y(i,i)=1, \quad 1\leq i,j\leq n.
\]
This optimization is an SDP relaxation of maximizing the Newman-Girvan modularity, and $\widehat{Y}$ is an estimate of $Y = \Pi \Pi'$. 
The method then obtains the first $K$ eigenvectors of $\widehat{Y}$ and applies a weighted $k$-median algorithm to $n$ rows of the $n\times K$ matrix of eigenvectors. 
\cite{cai2015robust} further developed the idea into a robust community detection algorithm for networks where some of the nodes are  adversarial outliers.

\cite{ma2020universal} proposed the CMM approach, where they model the network with a latent space 
model. Under this model, each node is associated with a latent feature vector $z_i\in\mathbb{R}^K$, 
and  $(A(i,j)|z_1,\ldots,z_n)  \sim  \mathrm{Bernoulli}(\mathrm{logit}(G(i,j)))$ with $G(i,j) =\alpha_i+\alpha_j+z_i'z_j$. 
For community detection, they first estimate $\{(\alpha_i,z_i)\}_{1\leq i\leq n}$ by a penalized MLE approach,  and then 
apply k-means to $\hat{z}_1, \hat{z}_2, \ldots, \hat{z}_n$.

Table~\ref{tab:error} compares SCORE and SCORE+  with LSCD, CMM, and the profile likelihood approaches  (tabu, PPL, DCPPL).   For the polblog data set, SCORE and SCORE+   have the lowest error rates. For the  Simmons and Caltech data sets,  SCORE+ outperforms CMM and LSCD, and CMM and LSCD outperform SCORE and the profile likelihood approaches.  
For the other $5$ data sets,  all methods perform similarly, except for (a) LSCD significantly underperforms for the football data, and (b) tabu and CMM slightly underperform  for the football and UKfaculty data, respectively.

\subsection{Hier-SCORE: SCORE for hierarchical network community detection} 
\label{subsec:HierSCORE} 
Communities in a network may have a tree structure. Let the whole network be the root.  We 
think the root consists of several communities,  each of which is further divisible, and so on 
and so forth. \cite{PhaseII-JBES} proposed Hierarchical-SCORE (Hier-SCORE) as an 
approach to hierarchical community detection, which combines SCORE and the 
cycle count statistic. Cycle count statistic is a recent idea for network global testing (e.g., \cite{GC,  gao2017testing, MaCycle, chang2020estimation}), where the main interest is to test whether the network has only one community (i.e., $K = 1$) or multiple communities (i.e., $K > 1$).  Among the recent works on global testing,  \cite{JKL2021} proposed the Signed-Quadrilateral (SgnQ) statistic as  
a special form of cycle count statistic, defined as follows. 
For $1 \leq i \leq n$, let $d_i$ be the degree of node $i$. Introduce a vector $\hat{\eta} \in \mathbb{R}^n$ by $\hat{\eta}_i = d_i / \sqrt{\sum_{j = 1}^n d_j}$. We center the adjacency matrix $A$ to $A^*$, where  $A^*(i, j) = A(i,j) - \hat{\eta}_i \hat{\eta}_j$, 
$1 \leq i\neq j \leq n$, and define the SgnQ test statistic by 
$Q_n = \sum_{i_1, i_2, i_3, i_4 (distinct)} A_{i_1 i_2}^*  A_{i_2 i_3}^* A_{i_3, i_4}^* A_{i_4 i_1}^*$.  
The SgnQ test statistic is then 
$\phi_n = [Q_n - 2 (\|\hat{\eta}\|^2 - 1)^2] / \sqrt{8 (\|\hat{\eta}\|^2 - 1)^4}$.    
\cite{JKL2021} showed that $\phi_n \goto N(0,1)$ if the null is true (i.e.,  $K = 1$), 
and $\phi_n$ is optimal in global testing for general DCMM models. 

The basic idea of Hier-SCORE is as follows. We first use SCORE to divide all nodes in the network to $K_0$ communities. Now, for each community, consider the sub-network by restricting nodes and edges to this 
community, and apply the SgnQ test $\phi_n$ and obtain an approximate $p$-value by $1 -\Phi(\phi_n)$ ($\Phi$ is the CDF of $N(0,1)$). If the $p$-value is larger than a threshold $\alpha_0$, we think the community is not 
further divisible. Otherwise, we further divide the community into several sub-communities (with a 
proper number of communities). We then continue with this until we decide none of the communities 
is  further dividable.  \cite{PhaseII-JBES} used the algorithm to build a 4-layer tree with 26 leaf communities for the 
co-authorship network of statisticians, where they set $\alpha_0 = 0.001$, and determine the number of 
communities in each layer by combining the scree plot and the  authors' partial knowledge of the statistics
community. See Section \ref{subsec:statistics} and \cite{PhaseII-JBES} for a more detailed discussion.

\subsection{Community detection by D-SCORE for directed networks} 
\label{subsec:DSCORE} 
In a directed network, the edges are directed, and the adjacency matrix $A$ is asymmetric. \cite{SCC-JiJin} 
proposed Directed-SCORE (D-SCORE) as an approach to community detection for directed networks. 
Suppose the network has $K$ communities. For $1 \leq k \leq K$, let $\hat{u}_k$ and $\hat{v}_k$ be the $k$-th left singular vector and $k$-th right singular vector of $A$, respectively. 
Let ${\cal N}_1$ and ${\cal N}_2$ be the support of $\hat{u}_1$ and $\hat{v}_1$, respectively.  
Fix a threshold $T > 0$ (similar to that of SCORE, we recommend $T= \log(n)$). Define $\widehat{U}, \widehat{V} \in  \mathbb{R}^{n, K-1}$ as follows where for $1 \leq k \leq K-1$ and $1 \leq i \leq n$,  
\[
\widehat{U}(i,k) = 
\left\{
\begin{array}{ll} 
\mathrm{sgn}(\hat{u}_{k+1}(i) / \hat{u}_1(i)) \min\{ |\hat{u}_{k+1}(i) / \hat{u}_1(i)|,T \}, &\mbox{if $i \in {\cal N}_1$},  \\ 
0, &\mbox{otherwise}  \\ 
\end{array}
\right., 
\]
and
\[
\widehat{V}(i,k) = 
\left\{
\begin{array}{ll} 
\mathrm{sgn}(\hat{v}_{k+1}(i) / \hat{v}_1(i)) \min\{ |\hat{v}_{k+1}(i) / \hat{v}_1(i)|, T\}, &\mbox{if $i \in {\cal N}_2$},  \\ 
0, &\mbox{otherwise}  \\ 
\end{array}
\right.. 
\] 
D-SCORE clusters all $n$ nodes into $K$ communities by applying  $k$-means to the $n$-rows of the matrix $[\widehat{U}, \widehat{V}] \in \mathbb{R}^{n, 2K-2}$.   In \cite{wang2020spectral}, the authors  proposed a Degree-Corrected Block Model for directed networks and provided theoretical guarantees for D-SCORE in a broad setting. 
They also proposed an improved version of D-SCORE which leads to better 
clustering results for nodes outside the set of ${\cal N}_1 \cap {\cal N}_2$. Recall that for a connected undirected  
network, by Perron's theorem \citep{HornJohnson}, we always have ${\cal N}_1 = {\cal N}_2 = \{1, 2, \ldots, n\}$. 
For a connected directed network (connectivity is defined by directed edges),  it may happen that the set ${\cal N}_1 \cap {\cal N}_2$ is smaller than $\{1, 2, \ldots, n\}$.

\subsection{Applications of SCORE, SCORE*,  D-SCORE, and Hier-SCORE to the networks of statisticians} \label{subsec:statistics}
In Section~\ref{subsec:twodata}, we present two data sets (Phase I and Phase II) on the publications of statisticians.  
\cite{SCC-JiJin} used the Phase I data to construct a co-authorship network and analyzed them with SCORE. 
Consider the co-authorship network where each author in the data set is a node (3,607 nodes in total) 
and two authors have an edge if and only if  they have co-authored at least one paper in the data range.  
\cite{SCC-JiJin} focused on the giant component of the network, which consists of 2,363 nodes. 
By combining the scree plot and their partial knowledge of the statistical community, 
the authors argued that the network consists of $3$ primary communities. By applying 
SCORE to the network with $K = 3$, they identified three communities which they interpreted as ``Bayes", ``Biostatistics", and ``High-Dimensional Data Analysis". It is possible that these communities are further divisible (e.g., by Hier-SCORE). 
 \cite{SCC-JiJin} also constructed a citation network using the Phase I data and analyzed it with D-SCORE.   The citation network has $2,654$ nodes,  where each node is an author and there is a directed edge from node $i$ to node $j$ if and only if  author $i$ has cited author $j$ at least once in the data range.   Using D-SCORE, the authors identified $3$ communities in the citation network, which they labeled as ``Large-Scale Multiple testing", ``Variable Selection", and ``Spatial and  Semi-parametric/Nonparametric Statistics". By their knowledge of the statistical community, 
the authors found that the first two are comparably easy to interpret, but the last 
one is hard to interpret, and the reason is that this community contains several 
sub-communities. The authors constructed a new network by restricting the nodes to the third community and 
identified $3$ sub-communities with D-SCORE. They found that the three sub-communities are much easier to interpret and labeled them as 
``Nonparametric Spatial / Bayes statistics",   ``Parametric Spatial Statistics", and ``Semiparametric/Nonparametric Statistics", respectively. 

Also using the Phase I data, \cite{li2020network} (see also \cite{li2020hierarchical})  analyzed a weighted citation network.    The sub-network has $706$ nodes (authors), where the weighted directed edge from node $i$ to node $j$ is the total number of citations from author $i$ to author $j$ in the data range. They identified 
$20$ research communities, each of which is a group of authors with similar research interests. Among them are  ``High-dimensional inference (sparse penalties)" and  ``Functional data analysis", etc. (the community names were manually assigned by them).

\begin{figure}[tb!]
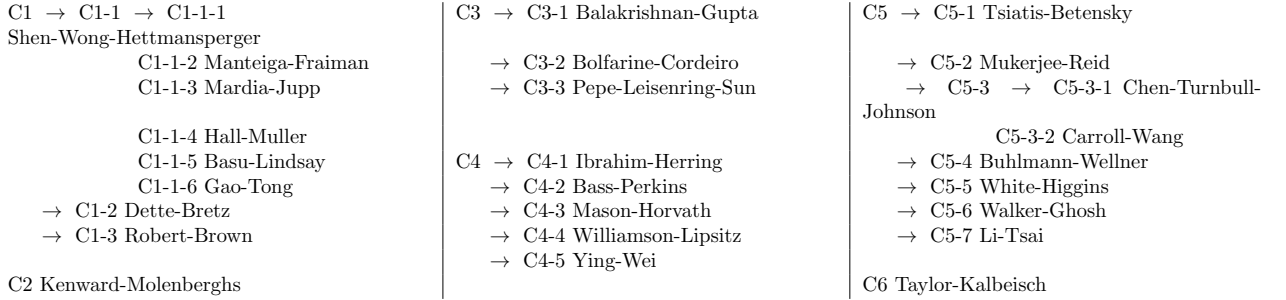
  
\centering
\scalebox{.78}{
\begin{tabular}{ >{\raggedright} p{7.2cm} | p{6.5cm}| p{6.8cm}}
C1 $\; \rightarrow\; $ C1-1\;  $\rightarrow$\; C1-1-1 Shen-Wong-Hettmansperger &  C3 $\; \rightarrow\; $  C3-1 Balakrishnan-Gupta & C5 $\;\rightarrow\; $ C5-1 Tsiatis-Betensky \\
\hspace{2.1cm}  C1-1-2 Manteiga-Fraiman &  \hspace{.37cm}  $\; \rightarrow\; $ C3-2 Bolfarine-Cordeiro   & \hspace{.37cm}  $\; \rightarrow\; $ C5-2 Mukerjee-Reid \\
\hspace{2.1cm} C1-1-3 Mardia-Jupp &  \hspace{.37cm}  $\; \rightarrow\; $  C3-3 Pepe-Leisenring-Sun   & \hspace{.37cm}  $\; \rightarrow\; $ C5-3 $\;\rightarrow\; $ C5-3-1 Chen-Turnbull-Johnson\\
\hspace{2.1cm} C1-1-4 Hall-Muller &   &  \hspace{2.15cm}  C5-3-2 Carroll-Wang\\
\hspace{2.1cm} C1-1-5  Basu-Lindsay &  C4 $\; \rightarrow\; $  C4-1 Ibrahim-Herring  &  \hspace{.37cm}  $\; \rightarrow\; $ C5-4 Buhlmann-Wellner \\
\hspace{2.1cm} C1-1-6  Gao-Tong & \hspace{.37cm}  $\; \rightarrow\; $ C4-2 Bass-Perkins & \hspace{.37cm}  $\; \rightarrow\; $ C5-5 White-Higgins \\
\hspace{.37cm} $\; \rightarrow\; $  C1-2 Dette-Bretz &     \hspace{.37cm}  $\; \rightarrow\; $ C4-3 Mason-Horvath   &  \hspace{.37cm}  $\; \rightarrow\; $ C5-6 Walker-Ghosh \\
\hspace{.37cm}  $\; \rightarrow\; $   C1-3 Robert-Brown &    \hspace{.37cm}  $\; \rightarrow\; $ C4-4 Williamson-Lipsitz  & \hspace{.37cm}  $\; \rightarrow\; $ C5-7 Li-Tsai  \\
 & \hspace{.37cm}  $\; \rightarrow\; $ C4-5 Ying-Wei & \\
C2 Kenward-Molenberghs & & C6 Taylor-Kalbeisch \\
\end{tabular}
}
\caption{The community tree obtained by SCORE on a coauthorship network of statisticians \citep{PhaseII-JBES}, where the network is constructed using published papers in 36 journals during 1975--2015.} \label{fig:community-tree} 
\end{figure}   

Using the Phase II data,  \cite{PhaseII-JBES} constructed a co-authorship network 
as follows.  We start with the network where each author in the data range is a node, and 
two nodes have an edge if and only if they have co-authored $k$ or more papers 
in the data range. If we choose $k = 1$ or $2$, the network is quite large and hard
to analyze, so \cite{PhaseII-JBES} took $k = 3$.  The network is 
comparably fragmented. We then focus on the giant component, which contains 4,383 nodes. 
We wish to identify interpretable communities in this network. As discussed in Section \ref{subsec:HierSCORE},   
large-scale networks of this kind may have hierarchical tree structures,  so an appropriate community detection approach is the Hier-SCORE introduced in Section \ref{subsec:HierSCORE}.  Note that in Hier-SCORE, we use SCORE each time we divide a community into several sub-communities.  As in Section \ref{subsec:SCORE*},   we may have an issue here, for some communities estimated this way may be dis-assortative, but a true co-authorship community is believed to be assortative. Fortunately, as in Section \ref{subsec:SCORE*},  such an issue can be nicely resolved if we replace SCORE by SCORE* each time when we divide a community to several sub-communities using Hier-SCORE. This gives rise to a small variant of Hier-SCORE, which we may call it Hier-SCORE*.  Using Hier-SCORE*,  \cite{PhaseII-JBES} built a $4$-layer community tree  for the co-authorship network above (the tuning parameters are set as in Section \ref{subsec:HierSCORE}). The tree has $26$ leaf communities, 
each of which is an interpretable community in statistics.  See Figure~\ref{fig:community-tree} (and also \cite[Figure 4]{PhaseII-JBES})  for the community tree,  where each leaf community is labeled by the last names of two or three authors selected by network centrality metrics.

\subsection{Tensor-SCORE for community detection with hypergraphs} 
\label{subsec:tensor} 
Different papers may have different numbers of authors.   To model the co-authorship relationship of all papers with 
exactly two authors, it is appropriate to use a network. However, to model the co-authorship of all papers with exactly $m$ authors for an $m \geq 3$, it is preferable to use an $m$-hypergraph. 
An $m$-hypergraph is captured by an order-$m$ supersymmetric tensor ${\cal A}$ (i.e., the adjacency tensor), where 
\[
{\cal A}(i_1, i_2, \ldots, i_m)  = \begin{cases}
1, & \mbox{if authors $i_1, \ldots, i_m$ have co-authored $\geq 1$ paper together},\\
0, &  \mbox{otherwise ($i_1, \ldots, i_m$ are distinct)}.  
\end{cases}
\] 
Community detection for $m$-hypergraphs is of great interest. It includes network community detection as a special case, but is more challenging. 
To extend SCORE to the hypergraph setting, one may replace the PCA part by an approximate Tucker decomposition for the adjacency tensor ${\cal A}$ and then apply SCORE on the resulting ``eigenvectors''  in a similar fashion. Some well-known algorithms for an approximate Tucker decomposition include HOSVD and HOOI \citep{kolda2009tensor}, but unfortunately, they suffered from either a slow rate of convergence or a potential `signal cancellation' (which is well-known in the literature of tensor analysis).  
\cite{ke2019community} proposed Tensor-SCORE as a  regularized power iteration algorithm for computing the approximate Tucker decomposition of the adjacency tensor $A$. The approach avoids `signal cancellation" and produces ``eigenvectors"  with a faster convergence rate, and overcomes the two challenges aforementioned.

Aside from hypergraph community detection,  the problem of hypergraph global testing is also of interest.  
In a broad hypergraph setting, 
 \cite{JKLiang2021} derived a sharp lower bound for global testing.  Also,  \cite{yuan2018testing} proposed an 
 interesting approach which can be viewed as a generalization of the cycle count idea for networks (e.g., 
 \cite{JKL2021}) to hypergraphs; see Sections  \ref{subsec:HierSCORE} and  \ref{subsec:EstK}.

\subsection{Estimating the number of communities $K$ by stepwise GoF, and the Non-Splitting Property (NSP) of SCORE}  \label{subsec:EstK}   
Most community detection methods need to know $K$ a priori. Therefore, 
how to estimate $K$ is a problem of great interest (e.g., \cite{FengEstK, ma2021determining, wang2017likelihood}).   As mentioned in Section \ref{subsec:HierSCORE}, 
the cycle count statistic (including the SgnQ test, which is optimal)  is a recent idea for network global testing \citep{JKL2021, GC, gao2017testing, MaCycle}.  \cite{jin2020estimating}  proposed stepwise GoF as a new approach to estimating $K$ by combining the idea of cycle count and SCORE, assuming that the network satisfies a DCBM model with $K$ (unknown) communities.  
The stepwise GoF runs as follows.   Input the adjacency matrix $A$ and $\alpha \in (0,1)$ (e.g., $\alpha = 5\%$). Let $z_{\alpha}$ be the upper-$\alpha$ percentile of $N(0,1)$. 
\begin{itemize}\itemsep0em 
\item For $m = 1, 2, \ldots$, apply SCORE assuming there are $m$ communities, and obtain the estimated community labels $\hat{\pi}_1^{(m)}, \ldots, \hat{\pi}_n^{(m)}$, where if node $i$ is assigned by SCORE to community $k$, then 
$\hat{\pi}_i^{(m)}(k) = 1$ and  $\hat{\pi}_i^{(m)}(\ell) = 0$ for $\ell\neq k$, $1 \leq k \leq m$, $1 \leq i \leq n$. 
\item Using $\hat{\pi}_1^{(m)}, \ldots, \hat{\pi}_n^{(m)}$ to fit with the adjacency matrix $A$ and obtain an estimate for $\Omega$, denoted by $\widehat{\Omega}^{(m)}$.  Let $\widetilde{A}^{(m)} = A - \widehat{\Omega}^{(m)}$. 
\item  Define  a cycle count statistic by $
Q_n^{(m)} = \sum_{i_1, i_2, i_3, i_4 (distinct)} \widetilde{A}_{i_1 i_2}^{(m)} \widetilde{A}^{(m)}_{i_2 i_3} 
\widetilde{A}^{(m)}_{i_3 i_4} \widetilde{A}_{i_4 i_1}^{(m)}$. Let $C_n =  \sum_{i_1, i_2, i_3, i_4 (distinct)} A_{i_1 i_2} A_{i_2i_3} A_{i_3i_4} A_{i_4 i_1}$ (i.e.,  total number of quadrilaterals in the network), and let $B_n^{(m)}$ be the estimated bias for $Q_n^{(m)}$.  Normalize 
$Q_n^{(m)}$ to  
$\psi_n^{(m)} = [Q_n^{(m)} - B_n^{(m)}] / \sqrt{8 C_n}$.  
\item Estimate $K$ by $\widehat{K}_n(\alpha) = \min\{m: \psi_n^{(m)} \leq z_{\alpha}\}$. 
\end{itemize} 
The expressions of  $\widehat{\Omega}^{(m)}$ and $B_n^{(m)}$ are a little bit long and so are omitted here. See 
\cite{jin2020estimating} for details. 
It was shown by \cite{jin2020estimating} that under some mild regularity conditions, 
\[
\mbox{$\psi_n^{(m)} \goto N(0,1)$ if $m = K$}, \qquad \mbox{and 
$\psi_n^{(m)} \goto \infty$ in probability if $m < K$}.  
\] 
If we let $\alpha$ depend on $n$ (i.e., $\alpha = \alpha_n$) and let $\alpha_n$ tend to $0$ slowly enough, 
then $P(\widehat{K}_n(\alpha_n) \neq K) \leq \alpha_n + o(1) \goto 0$, so $\widehat{K}_n(\alpha_n)$ is consistent for $K$.

To analyze $\psi_n^{(m)}$, we need to use concentration inequalities. However, 
the number of possible realizations for $[\hat{\pi}_1^{(m)}, \hat{\pi}_2^{(m)}, \ldots, \hat{\pi}_n^{(m)}]$ (i.e., the estimated community label matrix) may be as large as the order of $\mathrm{exp}(O(n))$, 
in which case the matrix  does not concentrate 
anywhere. \cite{jin2020estimating} overcame the challenge by proving that 
SCORE has the so-called {\it Non-Splitting Property (NSP)}: for any fixed $1 \leq m \leq K$, if 
we apply SCORE to an adjacency matrix $A$ generated from a DCBM with $K$-communities, then 
except for a small probability, each true community is contained in one of the communities estimated by SCORE (therefore, none of the true communities will be spilt into two or more parts, with different parts in different 
communities identified by SCORE). By NSP, the matrix $[\hat{\pi}_1^{(m)}, \hat{\pi}_2^{(m)}, \ldots, \hat{\pi}_n^{(m)}]$ 
concentrates on no more than $\mathrm{exp}(O(K))$ realizations (note that $n$ can be very 
large but $K$ is usually small;  see Table \ref{tab:degree}). Combining this with the union bound, we overcome 
the main technical hurdle for analyzing 
$\psi_n^{(m)}$; see  \cite{jin2020estimating}. 
The NSP of SCORE may be useful in many other network settings.

\subsection{SCORE for goodness of fit}  
For goodness of fit, we are interested in using the adjacency matrix to test whether the network is generated from a specified model.  An interesting version of the problem is to test the null hypothesis 
$H_0$:  $A$ is generated from a DCBM with $K$ communities (where $K$ is given).  
In this problem, the alternative is not specified, and can be either that the 
network has mixed-memberships so DCBM is inadequate, or that 
the network satisfies a DCBM with a different number of communities.  
The idea in  Section \ref{subsec:EstK} can be readily used to address this problem.  
In fact, $\psi_n = \psi_n^{(K)}$ is a good metric for goodness-of-fit, where under mild conditions (e.g., see \cite{jin2020estimating}), 
$\psi_n^{(K)}  \goto N(0,1)$ as $n \goto \infty$, when $K$ is correctly specified. 
 
The discussion above is for DCBM. 
The idea can also be used to test whether $A$ is generated from SBM with $K$ communities. 
Since SBM is a more idealized model (where we do not have severe degree heterogeneity) than DCBM,  the analysis is less challenging, and we may want to replace the SCORE step by OSC (since there is no need to remove the effects of degree heterogeneity).   
The idea can also be used to test whether $A$ is generated from a DCMM model with $K$ communities.  
This is a much more difficult case, as the network may both have severe degree heterogeneity and mixed-memberships. In this case, we first use Mixed-SCORE (to be introduced in Section \ref{sec:MME}) to estimate $\pi_i$'s. 
Once we have estimates for $\pi_1,\ldots,\pi_n$, we can use the similar idea as in Section~\ref{subsec:EstK} to obtain $\psi_n^{(K)}$, 
but the analysis of $\psi_n^{(K)}$ is expected to be much longer and more delicate.  See \cite{lei2016goodness} and \cite{hu2021using} for some recent approaches to goodness-of-fit. Also,  
in a related setting, \cite{SBMtesting} proposed to combine a degree-based chi-square statistic with the cycle count statistic for power enhancement in network global testing, and  \cite{arias2014community} used a degree-based chi-square statistic to test whether the network has a small planted clique.

\subsection{The exponential rate of SCORE and a sharp phase transition for perfect community detection} 
\label{subsec:SCORErate}   
Let $\widehat{\Pi}=[\hat{\pi}_1,\hat{\pi}_2,\ldots,\hat{\pi}_n]'$ be the matrix of estimated community labels by the orthodox SCORE. 
Define the Hamming error (per node)  for clustering  as $
\mathrm{Hamm}(\widehat{\Pi},\Pi) = n^{-1} \sum_{i=1}^n 1\{\hat{\pi}_i\neq \pi_i\}$, up to a permutation on columns of $\widehat{\Pi}$. 
\cite{SCORE} gave the first explicit error rate for $\mathrm{Hamm}(\widehat{\Pi},\Pi)$ under DCBM. \cite{SCORE+} further improved the error rate to an exponential form of $\theta$: Under some regularity conditions, they showed that 
\[
\mathbb{E}\big[\mathrm{Hamm}(\widehat{\Pi},\Pi) \big] \leq  \frac{2K}{n}\sum_{i=1}^n \exp\left( - a_2 \theta_i\cdot \min\biggl\{ \frac{(|\lambda_K|/\sqrt{\lambda_1})^2\|\theta\|^2}{K^2\|\theta\|_3^3},\; \frac{(|\lambda_K|/\sqrt{\lambda_1})\|\theta\|}{K\theta_{\max}}  \biggr\}\right)+o(n^{-3}), 
\]
where $a_2>0$ is a constant, $\theta_{\max}=\max_{1\leq i\leq n}\{ \theta_i\}$, and $\lambda_1,\ldots,\lambda_K$ are nonzero eigenvalues of $\Omega$ (arranged in the descending order of magnitude). 
\cite{gao2015achieving} and \cite{gao2016community} proposed methods that attain exponential error rates under a special ``assortative" DCBM, where the off-diagonal entries of $P$ are required to be strictly smaller than the diagonal entries of $P$. The results are very interesting, but whether they continue to hold for general DCBM (where the networks may be dis-assortative) remains unclear (note also it is hard to check whether a network 
is assortative in practice). 
Also, in the SBM setting (i.e., no degree heterogeneity), \cite{abbe2017entrywise} showed that the ordinary spectral clustering (OSC) attains an exponential error rate.  In comparison, the exponential rate of SCORE in \cite{SCORE+} is for more general settings, allowing for a rather arbitrary $P$ (community structure matrix) and severe degree heterogeneity. 
See Table~\ref{tab:error}, where we compare SCORE with the {\it refinement algorithm} in \cite{gao2016community} 
(abbreviated as {\it Refine}), which is essentially a combination of the idea of SCORE and {\it majority voting}. 
Refine slightly improves SCORE for the Simmons and Caltech data sets, but slightly underperforms SCORE for the polblog data set (the perform similarly for the other $5$ data sets). These suggest that 
majority voting may not help improve the error rates in general settings.  
 
To prove the exponential rate of SCORE, it requires sharp {\it row-wise} large deviation bounds for the matrix $\hat{\Xi}$ that contains the first $K$ eigenvectors of $A$. Specifically, we need to characterize the asymptotic behavior of $\|e_i'(\hat{\Xi}O-\Xi)\|$, for every $1\leq i\leq n$, where $e_i$ is the $i$th standard basis vector of $\mathbb{R}^n$ and $O\in\mathbb{R}^{K,K}$ is a (stochastic) orthogonal matrix that emerges in the eigen-decomposition of $A$. This is much more challenging than obtaining the large-deviation bound for $\|\hat{\Xi}O-\Xi\|_F$. The analysis requires sophisticated technical tools, which we review in Section~\ref{subsec:MSCORErate}.

Using the results on the exponential rates, we can deduce a sharp phase transition result for SCORE as follows.  
Consider the problem of {\it exact recovery} or {\it perfect community detection}: under what conditions, 
can we fully recover all community labels (so the Hamming error is $0$) with overwhelming probability? 
When $K$ is finite and $\theta_{\max}\leq C\theta_{\min}$,  it was shown in \cite{SCORE+} that SCORE achieves exact recovery if $|\lambda_K|/\sqrt{\lambda_1}\geq C$ for a sufficiently large constant $C>0$, where 
we recall $\lambda_k$ is the $k$-th largest (in magnitude) eigenvalue of $\Omega$ (and 
$\Omega$ only has $K$ nonzero eigenvalues).  At the same time, 
 \citep{JKL2021, jin2020estimating} showed that when $|\lambda_K|/\sqrt{\lambda_1}\to 0$, there exist two sequences of DCBM that have different $K$ but are asymptotically indistinguishable from each other. In this case, consistent estimates for $K$ do not exist and so exact clustering is impossible. Together, this shows that SCORE yields a sharp phase transition for the problem of exact recovery.  This also suggests that the condition of $|\lambda_K|/\sqrt{\lambda_1}\geq C$ is required 
 for exact recovery and can not be relaxed for general DCBM settings.

\section{SCORE normalization for mixed-membership estimation} \label{sec:MME}  
In Section \ref{sec:SCORE}, we focus on the DCBM model, where we 
allow severe degree heterogeneity but not mixed-memberships. For some networks (e.g., the co-authorship network in Section~\ref{subsec:statistics}), DCBM is appropriate, for the fraction of nodes with significant mixed-memberships is small. However, for some 
other networks (e.g.,  the citee network in Section \ref{subsec:citee}), many nodes may have significant mixed-memberships. In such a case, it is preferable to consider the DCMM model (see Section~\ref{subsec:DCMMmodel}) where we allow arbitrary mixed-memberships.  
In this section, we focus on the problem of estimating $\pi_1, \pi_2, \ldots, \pi_n$ (i.e., membership estimation). 
This includes the problem of community detection in Section \ref{sec:SCORE} as a special case.

\subsection{The ideal simplicial cone and ideal simplex for general DCMM, and the ideal Mixed-SCORE algorithm}  \label{subsec:idealsimplexDCMM} 
Section \ref{subsec:idealsimplexDCBM} discusses the ideal simplex for general DCBM, where each membership vector $\pi_i$ is degenerate (with only one nonzero entry).  
We now discuss the ideal simplex for the broader DCMM model, where each $\pi_i$ may be non-degenerate. 
Consider a DCMM model and let $\Omega = \Theta \Pi P \Pi' \Theta$ be the main signal matrix as in (\ref{DCMMmatrix}). 
As before, let $\lambda_k$ be the $k$-th largest (in magnitude) eigenvalue of $\Omega$, and let $\xi_k$ be the corresponding eigenvector.  Same as in Section \ref{subsec:idealsimplexDCBM}, let $\Xi = [\xi_1, \ldots, \xi_K] = [x_1, \ldots, x_n]'$; there exists a non-singular matrix $B = [b_1, \ldots, b_K]$ such that $\Xi=\Theta\Pi B$. Let $R = [r_1, \ldots, r_n]'$ and 
$V = [v_1, \ldots, v_K]$, with $R(i,m) = \xi_{m+1}(i)/\xi_1(i)$,  and 
$v_k (\ell) = b_{\ell+1 }(k) / b_1(k)$, $1 \leq i \leq n, 1 \leq m, \ell \leq K-1,  1 \leq k \leq K$. 
Denote the Hadamard product by $\circ$ and define the weight vectors $w_1, w_2, \ldots, w_n$ by 
\begin{equation} \label{Definew}  
w_i = (\pi_i \circ b_1) / \|\pi_i \circ b_1\|_1, \qquad 1 \leq i \leq n. 
\end{equation} 
Recall that node $i$ is a mixed node if $\pi_i$ is non-degenerate, and it is a pure node in community $k$ if $\pi_k(k) = 1$ and $\pi_k(\ell) = 0$ for $\ell \neq k$, $1 \leq i \leq n, 1 \leq k \leq K$. 
The following results are from \cite{MSCORE}, which extend our results on ideal simplex for DCBM to 
the broader DCMM, and give new insight we did not see before in Section \ref{subsec:idealsimplexDCBM}. 
\begin{lemma} \label{lemma:b1} 
For $1 \leq k \leq K$, $b_1(k) = [\lambda_1 + v_k' \diag(\lambda_2, \ldots, \lambda_K) v_k]^{-1/2}$.  
\end{lemma} 
\begin{theorem} \label{thm:idealcone} 
{\bf (Ideal simplicial cone for DCMM)}.  
The vectors $b_1, b_2, \ldots, b_K$ expand a 
simplicial cone in $\mathbb{R}^K$ by ${\cal S}_0 = \{\alpha_1 b_1 + \alpha_2 b_2 + \ldots \alpha_K b_K: \alpha_1, \alpha_2, \ldots, \alpha_K \geq 0\}$. 
The cone has $K$ edges, and the $k$-th edge is the ray defined by $\{\alpha_k b_k: \alpha_k \geq 0\}$. 
If we view each row of $\Xi$ as a point in $\mathbb{R}^K$, then row $i$ falls on one of the edges if 
node $i$ is pure, and falls in the interior of the cone otherwise, $1 \leq i \leq n$. 
\end{theorem} 
\begin{theorem} \label{thm:idealsimplex} 
{\bf (Ideal simplex for DCMM}).  There is a simplex ${\cal S}$ in $\mathbb{R}^{K-1}$ with $v_1, v_2, \ldots, v_K$ as the vertices.   If we view each row of $R$ as a point in $\mathbb{R}^{K-1}$, then row $i$ falls on one of the vertices if  
node $i$ is pure, and falls in the interior of the simplex otherwise, $1 \leq i \leq n$.  Moreover, we can express each row of $R$ as a convex linear combination of the $K$ vertices, 
\begin{equation} \label{findw} 
r_i = \sum_{k = 1}^K w_i(k) v_k, \qquad \mbox{where the $K$-dimensional weight vector $w_i$ is as in (\ref{Definew})}.  
\end{equation} 
\end{theorem} 
The weight vectors $w_1, \ldots, w_n$ are ``bridge quantities" and play a key role. By \eqref{findw}, we can retrieve $\pi_1, \pi_2, \ldots, \pi_n$ by {\it ideal Mixed-SCORE} below.  
\begin{itemize} \itemsep 0em  
\item Obtain the first $K$ eigenvalues $\lambda_1,  \ldots,  \lambda_K$ of $\Omega$ (arranged descendingly in magnitude), and obtain the corresponding eigenvectors $\xi_1, \ldots, \xi_K$. Define $R \in \mathbb{R}^{n, K-1}$ by $R(i,k) = \xi_{k+1}(i) / \xi_1(i)$, 
$1 \leq k \leq K-1$, $1 \leq i \leq n$. 
\item Identify the $K$ vertices by finding the convex hull of the rows of $R$ (there are many polynomial-time algorithms, e.g., \cite{preparata2012computational}). The recovered vertices are $v_1, v_2, \ldots, v_K$ (up to a permutation).   
\item Use $\{(\lambda_k, v_k)\}_{k = 1}^K$ to obtain $b_1$,  following Lemma~\ref{lemma:b1}.  For each $1 \leq i \leq n$, write $r_i$ as a convex combination of $v_1, v_2, \ldots, v_K$ as in Theorem \ref{thm:idealsimplex}. The weight vector in the combination coincides with $w_i$ in (\ref{Definew}).  
Define $\pi_i^* \in \mathbb{R}^K$ by $\pi_i^*(k) = 
w_i(k) / b_1(k)$.  We then have $\pi_i = \pi_i^* / \|\pi_i^*\|_1$. 
\end{itemize}

\subsection{Why other post-PCA normalizations may not work for membership estimation in general DCMM settings} 
\label{subsec:postPCADCMM} 
In Section~\ref{subsec:SCORE-type}, we have mentioned several other SCORE-type post-PCA normalizations. 
If the underlying model is DCBM and the goal is community detection, many SCORE-type normalizations may work nicely. 
Take SCORE$_1$ for example. In the ideal DCBM case, 
SCORE normalizes the $n \times K$ matrix $\Xi = [\xi_1, \xi_2, \ldots, \xi_k]$ to an $n \times (K-1)$ matrix $R$ as above, 
and SCORE$_1$ normalizes $\Xi$ to an $n \times K$ matrix, say, $R^*$. 
For either $R$ or $R^*$, we only have $K$ distinct rows, and all rows corresponding to the nodes in the same 
community are identical.  In such a case, with either $R$ or $R^*$, we can retrieve all community labels conveniently by applying the $k$-means to the $n$-rows of the matrix, so both normalization approaches work well. 

\begin{figure}[htb!]
\centering
\includegraphics[width = .25\textwidth, height=0.2 \textwidth]{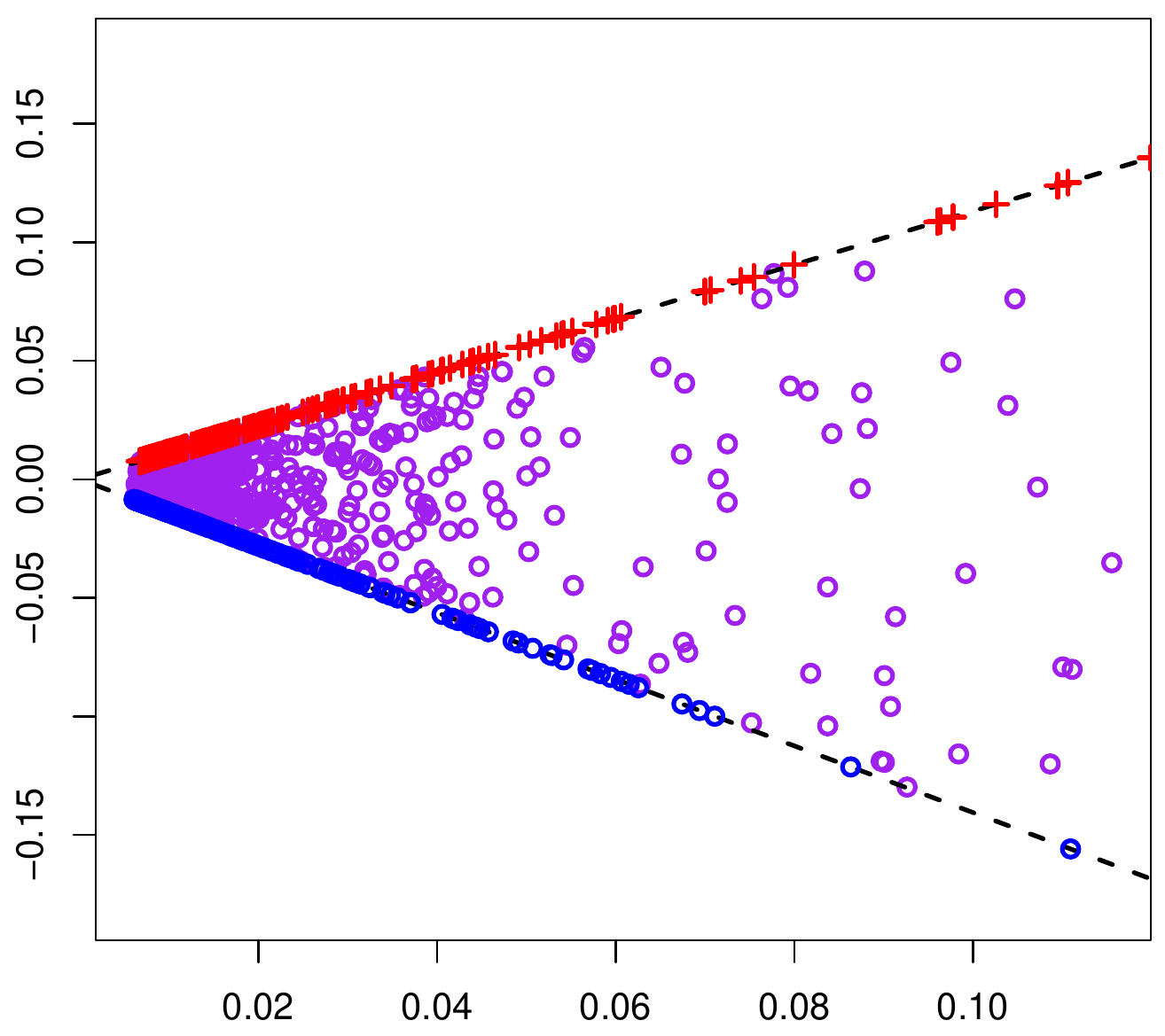} $\qquad$
\includegraphics[width = .25\textwidth,height=0.2\textwidth]{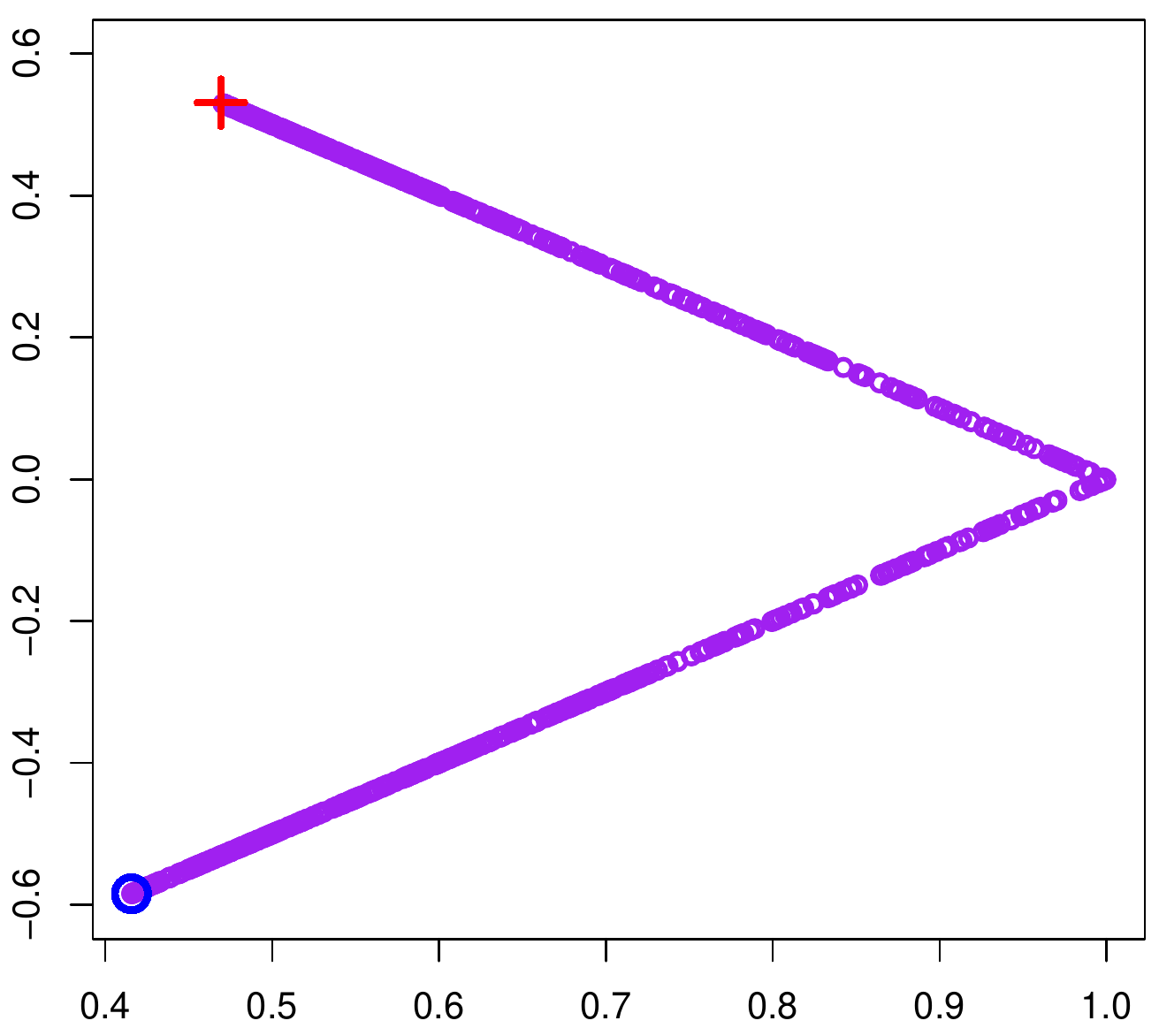}. $\qquad$
\includegraphics[width = .25\textwidth,height=0.2 \textwidth]{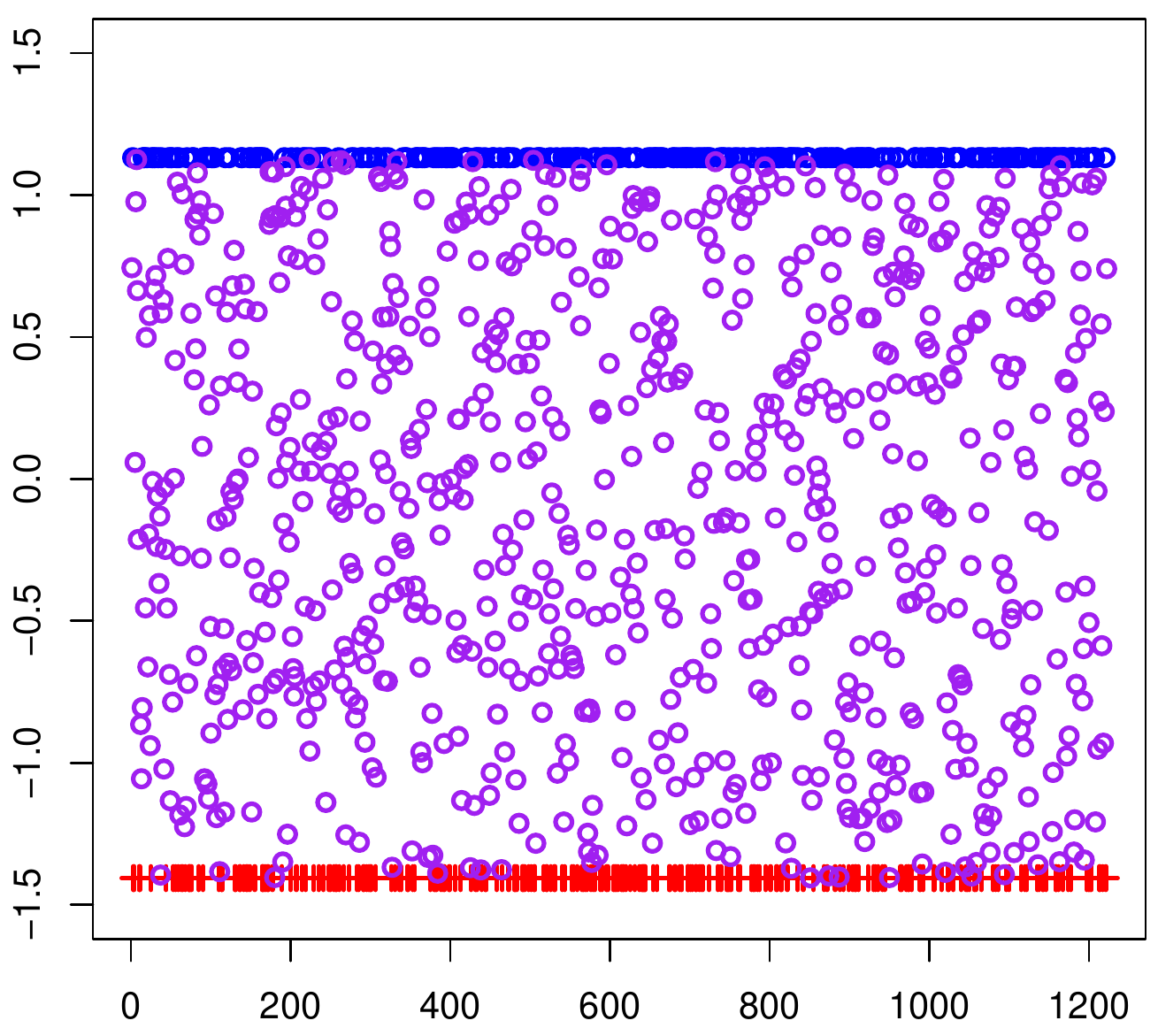}
\caption{Why other post-PCA normalization approaches do not work for membership estimation. Left: each point is a row $\Xi$. Middle: each point is a row of $R^*$. 
Right: each point is $(i, R(i))$, for $i = 1, 2, \ldots, n$; $\Xi$, $R$, and $R^*$ are 
as in Section \ref{subsec:postPCADCMM}.   
The network has two communities (i.e., $K = 2$). 
Red (or blue): a row for a pure node of community $1$ (or $2$). 
Purple: a row for a mixed-node.  
 There is a simplex underlying the rows of $R$ (corresponding to the SCORE normalization) 
 but there is no simplex underlying the rows of $R^*$ (corresponding to the SCORE$_1$ normalization). In the latter, 
 it is unclear how to define the ``bridge quantities" $w_i$ as in (\ref{Definew}) and so it is unclear how to retrieve $\pi_i$ as we do in the Ideal Mixed-SCORE.}    
\label{fig:other-normalization}
\end{figure}

However, the story is quite different if the underlying model is the much broader DCMM model and the goal is to estimate 
$\pi_1, \ldots, \pi_n$. Similarly, let us consider the ideal DCMM case.  With the SCORE normalization, first,  we can derive a  simple relationship between the ``bridge quantities" $w_i$ and $\pi_i$ as in (\ref{Definew}). Second, using the ideal simplex structure, we can 
find $w_i$ as in (\ref{findw}).  Together, this gives a convenient way to retrieve $\pi_i$ as in Ideal Mixed-SCORE.  Suppose we now choose to use the SCORE$_1$ normalization. Then first, it is unclear how to define the ``bridge quantities" $w_i$ and obtain a relationship between $w_i$ and $\pi_i$ similar to that of (\ref{Definew}). Second,  we no longer have an ideal simplex underlying the rows of $R^*$ (here, the matrices $R$ and $R^*$ are the same as above); therefore, even if we can define $w_i$ in some ways as desired, the approach in (\ref{findw}) is no longer valid,  and it is unclear how to retrieve $w_i$ using the matrix $R^*$. As a result, 
 the Ideal Mixed-SCORE approach above will not work if we replace the  SCORE normalization   by the SCORE$_1$ normalization.  This is illustrated in Figure~\ref{fig:other-normalization} where $(n, K) = (500,2)$ and $\theta^{-1}_1, \ldots,   \theta^{-1}_n$ are iid from $\mathrm{Uniform}([1,20])$. Also,  for the $2 \times 2$ matrix $P$, $P(i,j) = 1$ if $i = j$ and $P(i,j) = 0.1$ otherwise.  
Last, each of the two communities has $50$ pure nodes. The remaining $400$ are mixed nodes, where $\pi_i$ are iid from Dirichlet($1,1$).    In this case, for the $n$ rows of $R$, 
we have a simplex structure. For the $n$ rows of $R^*$, we no longer have a simplex structure: 
the $n$ rows fall on two line segments in $\mathbb{R}^2$, and it is unclear how to define the 
``bridge quantities" $w_i$.   By far,  SCORE is the only post-PCA normalization known to us that produces an ideal simplex and yields a convenient approach to membership estimation.

\subsection{Network membership estimation by Mixed-SCORE} 
The {\it Ideal Mixed-SCORE} is for the oracle case where $\Omega$ is accessible. 
We can extend the idea to the real case, but the main challenge is that, 
the ideal simplex is corrupted by noise. Therefore, we can no longer use a 
convex hull algorithm to retrieve the $K$ vertices of the ideal simplex.  
We need a more sophisticated algorithm (i.e., vertex hunting). 
Recall that $R = [r_1, r_2, \ldots, r_n]'$ and there is an ideal simplex underlying $r_1, r_2, \ldots, r_n$.  
\begin{definition} 
(Vertex Hunting (VH)). Suppose the convex hull of the $(K-1)$-dimensional vectors $r_1, \ldots, r_n$ is a simplex with $K$ vertices $v_1, v_2, \ldots, v_K$.  The vectors $r_1, \ldots, r_n$ and $v_1, \ldots, v_K$ are non-random and unknown, but we observe 
vectors $\hat{r}_1, \ldots, \hat{r}_n$, where $\hat{r}_i$ 
is a stochastic proxy of $r_i$, $1 \leq i \leq n$. A Vertex Hunting (VH)  
algorithm does the following job: it uses $\hat{r}_1,\hat{r}_2,\ldots,\hat{r}_n$ to produce an estimate for 
$v_1, v_2, \ldots, v_K$, denoted by $\hat{v}_1,\hat{v}_2,\ldots,\hat{v}_K$. 
\end{definition}
The Mixed-SCORE   by \cite{MSCORE} 
extended the ideal Mixed-SCORE to the real case, provided with a VH algorithm. It runs as follows. 

\begin{itemize} 
\itemsep0em 
\item {\it SCORE step}. Obtain the first $K$ eigenvalues of $A$ (arranged descendingly in magnitude), and let  $\hat{\xi}_1, \hat{\xi}_2, \ldots, \hat{\xi}_K$ be the corresponding eigenvectors. Let $\widehat{R} \in \mathbb{R}^{n, K-1}$ be the matrix $\widehat{R}(i,k)= \mathrm{sign}(\hat{\xi}_{k+1}(i)/\hat{\xi}_1(i)) \cdot \min\bigl\{|\hat{\xi}_{k+1}(i)/\hat{\xi}_1(i)|, \; T\bigr\}$. Write $\widehat{R} = [\hat{r}_1, \hat{r}_2, \ldots, \hat{r}_n]'$. 
\item Apply the VH algorithm to $\hat{r}_1,\hat{r}_2,\ldots,\hat{r}_n$ and obtain the estimated vertices $\hat{v}_1, \hat{v}_2, \ldots, \hat{v}_K$ (see Section~\ref{subsec:SVS} on VH algorithms). 
\item Obtain $\hat{b}_1$ by
$\hat{b}_1(k) = [\hat{\lambda}_1 + \hat{v}_k' \diag(\hat{\lambda}_2, \ldots, \hat{\lambda}_K) \hat{v}_k]^{-1/2}$, $1 \leq k \leq K$.  
For each $1 \leq i \leq n$, estimate $\pi_i$ as follows: Solve $\hat{w}_i\in\mathbb{R}^K$ from the linear equations: $\hat{r}_i = \sum_{k = 1}^K \hat{w}_i(k) \hat{v}_k$, $\sum_{k=1}^K\hat{w}_i(k)=1$. Obtain $\hat{\pi}_i^* \in \mathbb{R}^{K}$ by 
$\hat{\pi}_i^*(k) = \max\{0, \hat{w}_i(k)/\hat{b}_1(k)\}$, $1 \leq k \leq K$. Let $\hat{\pi}_i  = \hat{\pi}_i^*/\|\hat{\pi}_i^*\|_1$.  
\end{itemize}

\subsection{Vertex Hunting by Sketched Vertex Search (SVS), and the five variants of SVS}  
\label{subsec:SVS}  
Mixed-SCORE is a generic method, where for any vertex hunting approach (say, X),  we can plug it in 
and use Mixed-SCORE-X as an approach to membership estimation. \cite{MSCORE} introduced 
four different vertex hunting methods (SVS$^*$, SVS$_0$, CVS, and SP) umbrellaed by the name of {\it Sketched Vertex Search (SVS)}.  See Table \ref{tab:CompSVS},  where SVS$^+$ is a new 
variant in \cite{DYChen2} (see also \cite{VALISE}). 
SVS is motivated by the observations that in many network settings, the vectors $\hat{r}_1, \ldots, \hat{r}_n$ 
are relatively noisy, with many outliers, so it is preferable to reduce noise before we hunt for vertices. 
SVS is a two-stage algorithm. The first stage is a denoise stage, where we choose an integer $L \leq n$ and 
cluster all $n$ points into $L$ clusters by $k$-means.  
For each of the $L$ estimated clusters, we call the (Euclidean) center of the cluster the ``local center".   
By the nature of k-means, each local center is the average of a few nearby points and so more robust to outliers. Now, if   each community has a few pure nodes, then there are a  few points near each vertex, so that k-means will place at least one local center  near each vertex.   As a result, the first stage of the procedure produces $L$ possible candidates for the $K$ vertices, and largely reduces noise.  
In the second stage, among all $L$ ``local centers", we search for which $K$ of them are the best approximations to the $K$ vertices of the Ideal Simplex.  Here,  if $L$ is relatively small, we may use an exhaustive combinatorial search algorithm. Otherwise, we may use the {\it Successive Projection (SP)} algorithm \citep{araujo2001successive,nascimento2005vertex}. SP is a greedy algorithm. It is computationally more efficient than combinatorial search, but it may be susceptible to noise and vulnerable to outliers. Therefore, it is preferable to combine SP with the denoise stage than directly using SP for vertex hunting.  
Depending on what $L$ we choose in the first stage and what algorithm we choose in the second stage, SVS has $4$ variants as in Table \ref{tab:CompSVS}: SVS$_0$, SVS*, CVS, and SP. Here,  CVS standards for {\it Combinatorial Vertex Search}, 
and SVS$_0$ is the original SVS proposed by \cite{MSCORE}. In SVS$_0$ and SVS*, it is desirable to select $L$ in a data-driven fashion, where we can use the procedure proposed in \cite{MSCORE} (Equation (1.13)).  
\begin{table}[hbt!] 
\centering 
\scalebox{0.9}{ 
\begin{tabular}{ccccc}
 & Skipping the 1st stage & Using $k$-means in 1st stage & Using KNN in 1st stage\\
 \hline
Using exhaustive search in 2nd stage & CVS & SVS$_0$ & --\\
Using SP in 2nd stage  & SP & SVS* & SVS$^+$\\
\hline 
\end{tabular}
} 
\caption{Comparison of $5$ versions of SVS. In SVS$_0$ and SVS*, we 
can use the data-driven choice of $L$ in Equation (1.13) of \cite{MSCORE}. 
SVS$^+$ has two tuning parameters $(m, N)$. The algorithm is relatively insensitive 
to tuning, and we usually set $(m, N) =(3, n/10)$ (on a high level, we may view SVS$^+$ as a variant of SVS with some $L < n$,  but there is no direct relationship between $(m, N)$ and $L$).}  \label{tab:CompSVS}
\end{table}

In \cite{DYChen2} (see also \cite{VALISE}), the authors proposed a new algorithm 
which we may call SVS$^+$.  The method is similar to SVS*, but instead of 
using $k$-means clustering to denoise in the first stage, we use 
the classical $k$-nearest neighborhood (KNN) (in theory, $k$-means may have a much higher 
computational cost than that of KNN; however, in practice, one usually uses 
the well-known Lloyd's algorithm \citep{lloyd1982least}  to implement the $k$-means clustering,  which runs reasonably fast).  SVS$^+$ runs as follows. 
\begin{itemize} \itemsep 0em
\item {\it Denoise by K-nearest neighbors (KNN)}. 
Let $\ell_{\max}=\max_{i\neq j}\|\hat{r}_i-\hat{r}_j\|$. For each $\hat{r}_i$, count the total number of $\hat{r}_j$'s that fall within a distance of $0.05\ell_{\max}$; if this number is strictly smaller than $m$, remove $\hat{r}_i$; otherwise, replace $\hat{r}_i$ by $\bar{r}_i$, where $\bar{r}_i$ is the average of the $N$-nearest neighbors (including $\hat{r}_i$ itself) of $\hat{r}_i$. Denote by ${\cal S}$ the index set of remained data points. This gives rise to the post-denoise data cloud  $\{\bar{r}_i: i\in {\cal S}\}$. 
\item {\it Vertex search by successive projection (SP)}. Apply the SP algorithm on the post-denoise data cloud to obtain $\hat{v}_1,\hat{v}_2,\ldots,\hat{v}_K$.
\end{itemize}
We also include SVS$^+$ in Table \ref{tab:CompSVS} for comparison.  
SVS$^+$ has two tuning parameters, the threshold $m$, and the number of nearest neighbors $N$. Our simulations suggest that the performance of SVS$^+$ is insensitive to the choice of $(m,N)$.  On a high level, we may view SVS$^+$ as a variant of 
SVS for some $L < n$, but there is no direct relationship between $(m, N)$ and $L$.

\begin{figure}[tb!]
\centering
\includegraphics[width=0.3\textwidth, height=.245\textwidth]{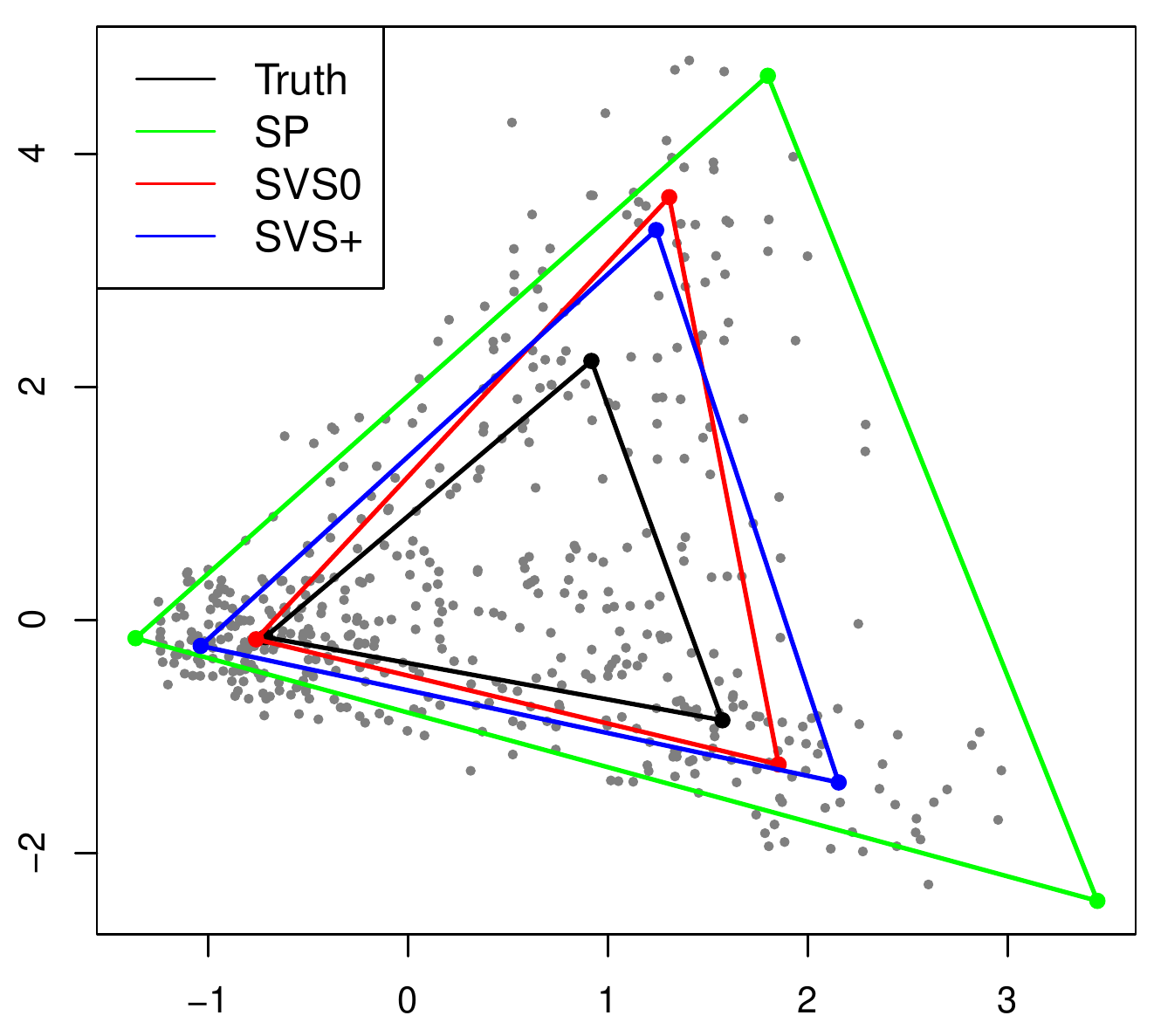}
\includegraphics[width=0.29\textwidth, height=.24\textwidth]{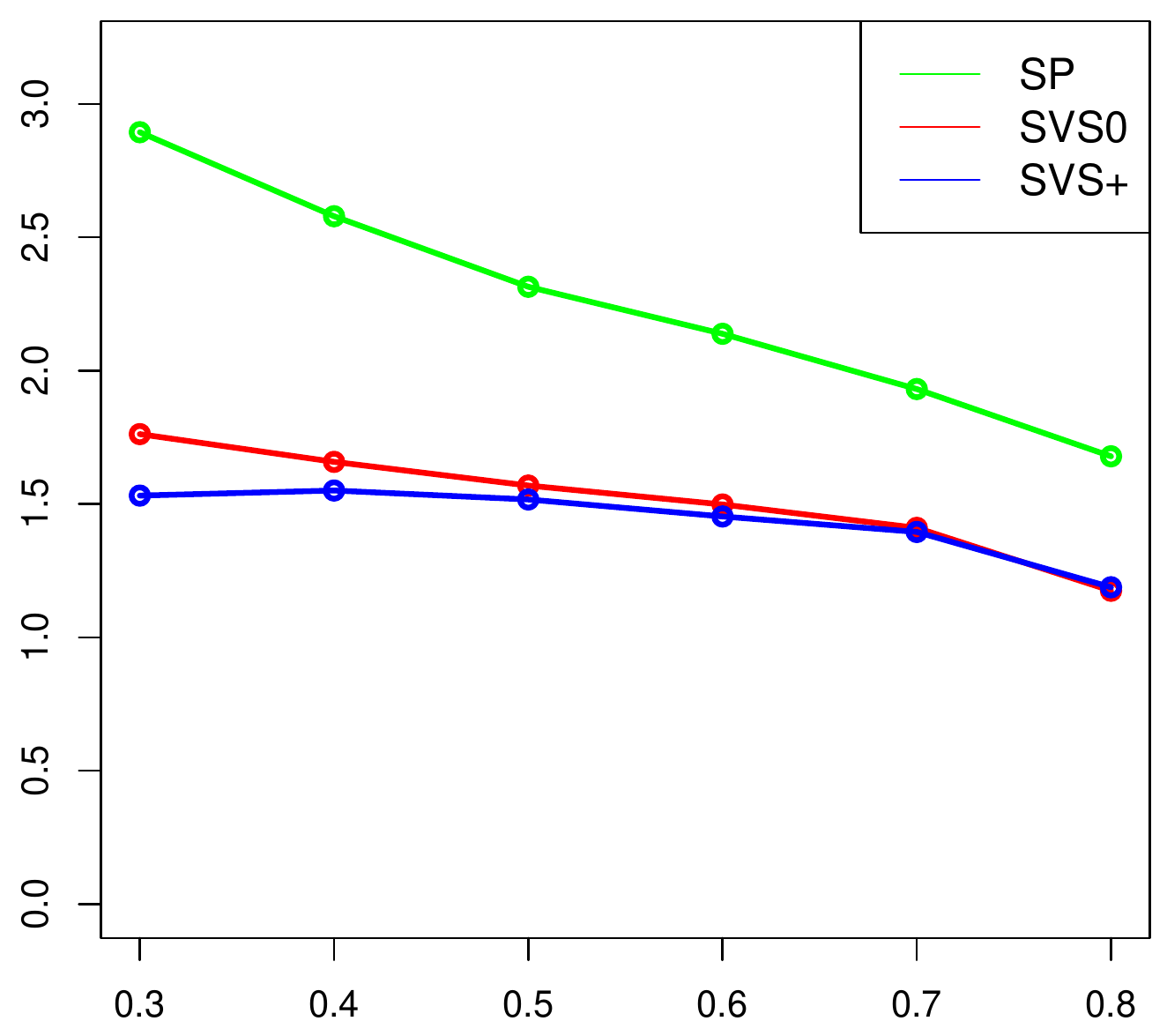}
\includegraphics[width=0.29\textwidth, height=.24\textwidth]{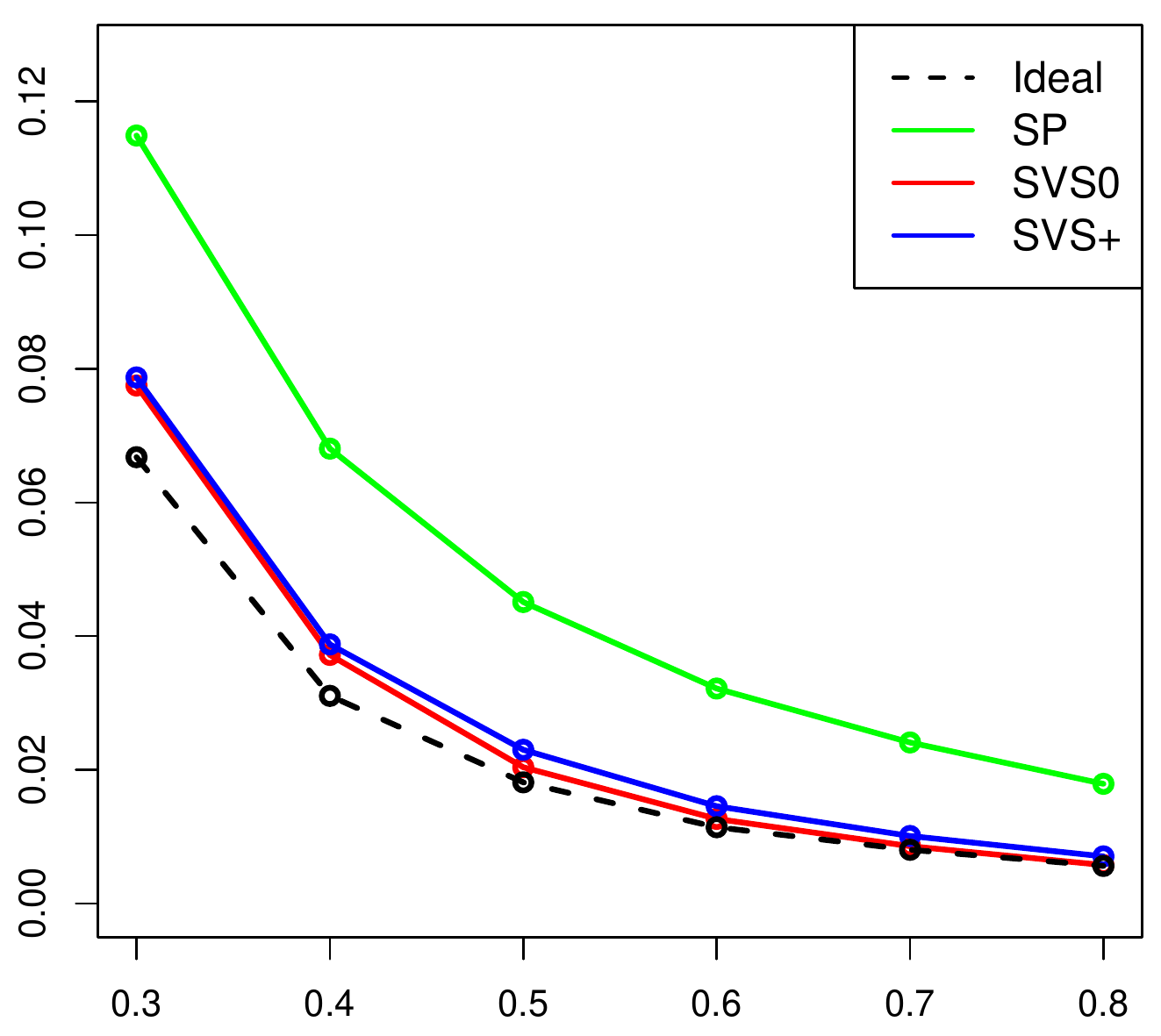}
\caption{Comparison of SP, SVS$_0$, and SVS$^+$. The simulation setting is as in Section \ref{subsec:SVS}, 
where $K = 3$, so the ideal simplex is a triangle. Left: the true ideal simplex (black) and 
the three estimated simplexes. Middle: the vertex hunting error $\max_{1\leq k\leq K}\|\hat{v}_k-v_k\|$ (over 50 repetitions and up to a permutation of the estimated vertices; x-axis is $\beta$, capturing the network sparsity). Right: the membership estimation error $\frac{1}{n}\sum_{i=1}^n\|\hat{\pi}_i-\pi_i\|^2$ (over 50 repetitions; x-axis is the same as above) by plugging the algorithms into Mixed-SCORE, where in the `Ideal' approach we plug in the true simplex. The results suggest 
that directly using SP for vertex hunting without a denoise stage may significantly underperform.  By comparing all $5$ variants of SVS in three aspects (theory, computation, and numerical accuracy), we recommend 
SVS$^+$ and SVS$^*$.} \label{fig:VH}
\end{figure}

We now compare the 5 variants of SVS in three aspects: theoretical convergence rate, computation,  
and numerical performance.  First,  \cite{MSCORE} showed that the convergence rates of SVS$_0$ and SVS* in  mean squared error (MSE) are faster than that of CVS and SP (SVS$^+$ is similar to SVS* and is expected to have a similar convergence rate); see Section \ref{subsec:MSCORErate} for more discussion.  
Second, the (theoretical) computational costs of CVS and SVS$_0$ are higher than those of 
SVS* and SP, respectively,  so we may prefer to use SVS* or SP when $(K, L)$ are relatively large (though CVS and SVS$_0$ may be more accurate when $(K, L)$ are relatively small, in which case computation is not an issue).   
The main computational cost of SVS* comes from the denoise stage where we use $k$-means 
clustering, which is high when $L$ is large. However, the real computing time of SVS* is reasonable if we use the well-known Llyod's algorithm. Also, when SVS* faces a computational challenge, we may 
choose to use SVS$^+$, which has a quite similar performance as that of SVS*, but 
has a polynomial-time computational cost.  Finally, numerically, SVS* and SVS$^+$ are much more accurate than SP. See Figure \ref{fig:VH}, where we present a simulation example with $(n, K)=(500, 3)$. The diagonals of $P$ are $1$ and off-diagonals are $0.1$; $\Pi$ has $50$ pure rows for each community, $175$ rows drawn from $\mathrm{Dirichlet}(\alpha_1)$ and $175$ rows from $\mathrm{Dirichlet}(\alpha_2)$, where $\alpha_1=(0.6, 0.2, 0.2)'$ and $\alpha_2=(0.3, 0.4, 0.3)'$; $\theta_i$'s are all equal to $\beta$. Here, $\beta$ captures the sparsity level, and we consider $\beta\in\{0.3, 0.4, \ldots, 0.8\}$; for each setting, we run 50 repetitions and report the average vertex hunting error and the average membership estimation error (by plugging the VH algorithm into Mixed-SCORE). The above results support the main point of SVS: the vectors $\hat{r}_1, \ldots, \hat{r}_n$ 
are relatively noisy with many outliers, so it is preferable to use a denoise stage before we hunt for vertices of the ideal simplex. For these reasons, we recommend SVS* and SVS$^+$ for practical use, especially when $(K, L)$ are relatively large. See \cite{winter1999n} and \cite{craig1994minimum} for other possible ideas for vertex hunting. 

\subsection{Some related works on membership estimation} 
\label{subsec:otherME} 
\cite{JiZhuMM} proposed a spectral method called OCCAM for mixed membership estimation, which runs as follows. 
First,  it applies SCORE$_2$ to $\widehat{\Xi}$ to get low-dimensional vectors $\hat{s}_1,\hat{s}_2,\ldots,\hat{s}_n$ on the unit sphere of $\mathbb{R}^K$. Second, it applies k-means to $\hat{s}_1,\hat{s}_2,\ldots,\hat{s}_n$, assuming $\leq K$ clusters.  Last, it uses the $K$ cluster centers output by k-means to construct estimates of $\pi_1,\pi_2,\ldots,\pi_n$.  Compared with Mixed-SCORE, 
there are several differences. First, OCCAM chooses to use SCORE$_2$ instead of SCORE for post-PCA normalization. 
As explained in Section \ref{subsec:postPCADCMM}, such a post-PCA normalization does  not necessarily produce a 
simplex, and it is unclear how to define the ``bridge quantities" $w_i$ (which is the main reason why Ideal Mixed-SCORE 
is able to retrieve all $\pi_i$ exactly).   As a result, OCCAM needs  stronger conditions to ensure consistency: for example,  they require the fraction of mixed nodes to be properly small,  while Mixed-SCORE does not need such a condition.  OCCAM may also have a convergence rate slower than Mixed-SCORE; see Table 2 of \cite{MSCORE}. 
The authors also propose OCCAM as a community detection approach. 
See Table \ref{tab:error}, where we compare SCORE with $15$  approaches including OCCAM.

\cite{airoldi2009mixed} proposed a method for mixed membership estimation under MMSBM, a special case of DCMM with $\theta_1=\ldots = \theta_n$ (e.g., see Section \ref{subsec:DCMMmodel} and Figure \ref{fig:model}). 
This is a Bayesian approach, where they imposed an iid Dirichlet prior on $\pi_1,\pi_2,\ldots,\pi_n$ and used a variational EM algorithm for estimating model parameters  
and a Markov Chain Monte Carlo algorithm for computing the posterior of $\pi_i$'s. 
Despite the practical popularity of the approach, 
the paper focused on the more idealized MMSBM setting and did not provide explicit theoretical guarantees.

In a closely related setting, \cite{fan2019simple} considered the multiple testing problem with the DCMM model, where for each pair of $1 \leq i\neq j \leq n$, we test $H_0: \pi_i=\pi_j$ versus $H_1: \pi_i\neq \pi_j$.  The result is useful (for example) in  constructing diversified portfolios in finance \citep{fan2019simple}.  They first applied SCORE to obtain $\hat{r}_i$ and $\hat{r}_j$ (same as above) and then constructed the test statistic $G_{ij}=(\hat{r}_i-\hat{r}_j)'\hat{\Sigma}^{-1}(\hat{r}_i-\hat{r}_j)$, where $\hat{\Sigma}\in\mathbb{R}^{K-1,K-1}$ is an estimate of the asymptotic covariance matrix of $\hat{r}_i-\hat{r}_j$. Under mild regularity conditions, they showed that $G_{ij}\to \chi^2_{K-1}$ under $H_0$ and that $G_{ij}\to\infty$ in probability under $H_1$. 
For multiple testing, we combine the test statistics for different pairs of $(i,j)$.

\subsection{Membership estimation by Dynamic Mixed-SCORE for dynamic networks}  
\label{subsec:dynamic-MSCORE} 
\cite{PhaseII-JBES} considered a dynamic network setting, where 
they extended DCMM for static networks to dynamic  
DCMM. Consider $T$ undirected networks for the same set of $n$ nodes, 
each for a time point $t$, $1 \leq t \leq T$. Let $A_1, A_2,   \ldots, A_T$ be the corresponding adjacency matrices.  
For each $1 \leq t \leq T$, we assume $A_t$ is generated from a DCMM model as in (\ref{DCMMmatrix}),  
with matrices $(\Theta^{(t)}, \Pi^{(t)}, P^{(t)})$.  We allow the matrices $(\Theta^{(t)}, \Pi^{(t)})$ to vary with time, but assume that the community structure matrix $P^{(t)}$ remains the same over time. We assume $A_1, A_2, \ldots, A_T$ are independent of each other given $\{(\Theta^{(t)}, \Pi^{(t)}\}_{t=1}^T$ (but such an assumption can be relaxed). Methods for dynamic mixed membership estimation were proposed by  
\cite{kim2018review, liu2018global}, but these works focused on 
a more restricted setting where each $A_t$ satisfies an MMSBM, which does not model degree heterogeneity (e.g., 
see Section \ref{subsec:DCMMmodel} and Figure \ref{fig:model}).  Under this dynamic DCMM, we may simply apply Mixed-SCORE to each $A_t$, but this may have unsatisfactory results. One challenge is that, 
the simplex associated with each $A_t$ depends on $(\Theta^{(t)}, \Pi^{(t)}, P)$, and these simplexes are not aligned properly with each other and may vary with time. 
As a result, it is hard to apply a smoothing scheme over different $t$ and borrow information across different networks.   
To fix the problem, \cite{PhaseII-JBES} proposed {\it dynamic Mixed-SCORE} (also, see \cite{dynamic-MSCORE}), which uses the ideal simplex associated with $A_1$ as a reference, and aligns the ideal simplexes associated with other $A_t$ with the reference simplex by 
a transformation. The algorithm runs as follows. 
Let $\hat{\lambda}_1,\hat{\lambda}_2,\ldots,\hat{\lambda}_K$ be the $K$ largest eigenvalues (in magnitude) of $A_1$, and let $\hat{\xi}_1, \hat{\xi}_2, \ldots, \hat{\xi}_K$ be the corresponding eigenvectors. For each $1\leq t\leq T$ and each node $1\leq i\leq n$, define a $(K-1)$-dimensional vector $\hat{r}_i^{(t)}$ by
\beq \label{dynamic-SCORE}
\hat{r}_i^{(t)}(k) = \frac{\hat{\lambda}_1(e_i' A_t \hat{\xi}_{k+1})}{\hat{\lambda}_{k+1}(e_i' A_t \hat{\xi}_1)},\qquad 1\leq k\leq K-1.   \qquad (\mbox{$e_i$: the $i$th standard basis vector of $\mathbb{R}^n$}). 
\eeq
\cite{PhaseII-JBES} showed that each $\hat{r}_i^{(t)}$ is a proxy of a non-stochastic vector $r_i^{(t)}$, where for each $1\leq t\leq T$, there is an ideal simplex ${\cal S}_t$ such that the point cloud $r_1^{(t)}, r_2^{(t)},\ldots,r_n^{(t)}$ are contained in ${\cal S}_t$ and that $r_i^{(t)}$ falls on one vertex of this simplex if and only if $i$ is a pure node at time $t$. 
The simplex ${\cal S}_1$ is the same as the one from applying Mixed-SCORE to $A_1$, but the other simplexes, ${\cal S}_2,\ldots,{\cal S}_T$, are different from the ones from Mixed-SCORE. Since all $T$ simplexes are obtained using the eigen-pairs of $A_1$, they are always properly aligned. This resolves the alignment issue aforementioned. Given these $\hat{r}_i^{(t)}$, we can similarly estimate the membership vectors by vertex hunting, which gives rise to the {\it dynamic Mixed-SCORE} algorithm \citep{dynamic-MSCORE}. See details therein.

\subsection{Applications of Mixed-SCORE and Dynamic Mixed-SCORE to the networks of statisticians} 
\label{subsec:citee} 
We first discuss an application of Mixed-SCORE to a co-authorship network. In Section \ref{subsec:statistics}, we mention a network with 2,263 nodes, which was constructed using the Phase I data  and was analyzed by \cite{SCC-JiJin}.  
The network is relatively large, and it is hard 
to identify tight-knit small research groups in statistics, so \cite{SCC-JiJin}  
considered a slightly different network   where each 
author in the Phase I data  is a node, but two nodes have an edge if and only if they have 
co-authored two or more papers in the data range. 
Compared to the previous co-authorship network, this network is 
more fragmental and reveals a number  
of tight-knit small-size research groups in statistics that can be interpreted in a meaningful way.   
The giant component of this network (with a total of 236 nodes)  is especially interesting, consisting of 
many co-authors of  Jianqing Fan (so we may call it the Fan's group for convenience,   though whatever name we use here, it is hard to represent all authors in the group). Not aware that the 
network may have significant mixed-memberships, 
\cite{SCC-JiJin} modeled the network with a DCBM (e.g., see Section \ref{subsec:DCMMmodel} and Figure \ref{fig:model})  and applied $4$ community detection approaches including SCORE.  All the methods agree that there are two communities which can be interpreted as 
the ``Carroll-Hall" community on semi-parametric  and non-parametric statistics, and 
the ``North-Carolina" community,  but their clustering results are very different. The reason is that (as pointed out later by \cite{MSCORE})  
many authors (e.g., Runze Li, Chunming Zhang) in the Fan's group have strong ties to both 
communities and thus have mixed-memberships. 
In such a case, community detection is not a proper problem to ask,  
and it is preferable to consider the problem of membership estimation, 
under the broader DCMM model.  Following this path,  \cite{MSCORE} applied 
Mixed-SCORE to this network, and obtained results that are more meaningful than those in \cite{SCC-JiJin}. 
In Table~\ref{tb:Fan-group}, we present the estimated memberships for 17 high-degree authors in the Fan's group (collaborators of Jianqing Fan); this table is  from \cite{MSCORE}. 

\begin{table}[htb!]  
\hspace*{-5em}
\scalebox{.82}{
\begin{tabular}{lc | lc | lc  | lc}
\hline
Name &  Membership & Name &  Membership & Name &  Estimated PMF & Name & Estimated PMF \\
\hline
Jianqing Fan  & 54\% of CH  & Yufeng Liu & 52\% of NC & Wenyang Zhang  & 51\% of CH & Per Aslak Mykland  &  52\% of NC \\
Jason P Fine  & 54\% of CH & Xiaotong Shen & 55\% of NC & Howell Tong 	 & 52\% of NC & Bee Leng Lee  &  54\% of CH \\
Michael R Kosorok	  & 57\% of CH & Kung-Sik Chan &  55\% of NC & Chunming Zhang  &  51\% of CH \\
J S Marron  & 55\% of NC & Yichao Wu &  51\%  of CH & Yingying Fan & 52\% of NC\\ 
Hao Helen Zhang & 51\% of NC & Yacine Ait-Sahalia & 51\%  of CH & Rui Song	 & 52\% of CH \\
\hline \\
\end{tabular}}
\caption{Estimated memberships for 17 high-degree authors in the Fan's group \cite{MSCORE}.   'CH' is short for `Carroll-Hall', and `NC' is short for `North Carolina'. Since $K = 2$, each estimated membership vector $\hat{\pi}_i$ only has two entries, and 
$54\%$ in `CH' means $46\%$ in `NC'.} \label{tb:Fan-group}
\end{table}

We now discuss an application of Dynamic Mixed-SCORE.  Using the Phase II data in Table \ref{tab:twodata}, 
\cite{PhaseII-JBES} constructed 21 citee networks, each for a time window  between 1990 and 2015 given in \cite[Table 1]{PhaseII-JBES}. In the citee network, each node is an author and two authors have an edge if and only if they have been co-cited at least twice  
by another author in the same time window. 
\begin{itemize} \itemsep0em
\item By applying Mixed-SCORE to the first network (time window 1991-2000), the authors discovered a statistical triangle, with the vertices representing the three primary research areas in statistics: ``Bayes", ``Biostatistics", and ``non-parametric statistics". The triangle is reminiscent of Efron's triangle of statistical philosophy \cite{Efron-Fisher}, but the latter is not based on real data. 
\item The authors further constructed the research map of the statisticians in the same 
two-dimensional plane of the statistical triangle. On the research map, each author who has published in the time window 
is represented as a point, and the relative position of the point to the three vertices represents the research interest or impact of the author. 
\item Using the first network (1991-2000) as the reference network,  \cite{PhaseII-JBES} 
used Dynamic Mixed-SCORE to develop an approach for computing the 
research trajectory for an author of interest in the time window.  The trajectory is a curve 
connecting the 21 points on the research map, each representing the estimated 
research interest of the author in a time window. See Table \ref{tab:trajectory}. 
\end{itemize} 
These results shed light on how research interest and impact of an individual author evolve over time and help us understand the research patterns and citation behavior of statisticians. 
\begin{table}[htb!]  
\centering
\scalebox{.8}{
\begin{tabular}{l cccccccc | l cccccccc}
\hline
\multicolumn{9}{c|}{Robert Tibshirani} & \multicolumn{9}{c}{Bin Yu}\\ 
\hline
time window  & 1 & 4 & 7 & 10 & 13 & 16 & 19 & 21 & time window  & 1 & 4 & 7 & 10 & 13 & 16 & 19 & 21  \\
\hline
Bayes & 3.30 & 3.19 & 3.04 & 2.97 & 2.92 & 2.82 & 2.90 & 2.83 & Bayes & 1.05 & 1.04 & 1.41 & 2.07 & 2.50 & 2.87 & 2.95 & 3.10 \\
Biostat. & 3.35 & 3.36 & 3.53 & 3.63 & 3.68 & 3.76 & 3.68 & 3.88 & Biostat. & 5.18 & 5.10 & 5.02 & 4.44 & 4.22 & 4.04 & 4.06 & 4.17 \\
Nonparam. & 3.05 & 3.15 & 3.12 & 3.09 & 3.08 & 3.11 & 3.11 & 2.99 & Nonparam. & 4.21 & 4.26 & 3.85 & 3.40 & 3.06 & 2.81 & 2.71 & 2.49 \\
\hline 
\end{tabular}
}
\caption{For author $i$, the trajectory is the curve connecting $\hat{r}_{i}^{(1)}, \ldots,\hat{r}_i^{(21)}$; see Sections~\ref{subsec:dynamic-MSCORE}~\ref{subsec:citee}. For $i = `Yu'$ and $i = `Tibshirani'$,  
Table \ref{tab:trajectory} reports the distance between $\hat{r}_i^{(t)}$ and each of the $3$ vertices of the Statistics Triangle (`Bayes', `Biostastistics' and  `Nonparametric statistics').   For an author,  if the distance to a vertex decreases over time, then his/her weight for the corresponding primary area increases. The table shows that in 1991-2015,  Yu's interest or impact changes significantly in 1991-2015, while that of Tibshirani changes less significantly; see Figures 1-2 of \cite{PhaseII-JBES}.} \label{tab:trajectory}
\end{table}

\subsection{Optimality of Mixed-SCORE, and comparison of rates for Mixed-SCORE-SP and Mixed-SCORE-SVS$^*$} \label{subsec:MSCORErate} 
Let $\hat{\Pi}=[\hat{\pi}_1,\hat{\pi}_2,\ldots,\hat{\pi}_n]'$ be the estimate by Mixed-SCORE. Define the mean-squared error as ${\cal L}(\hat{\Pi},\Pi)=\frac{1}{n}\sum_{i=1}^n\|\hat{\pi}_i-\pi_i\|^2$, up to a permutation of the columns of $\hat{\Pi}$.  \cite{MSCORE} derived the first explicit rate of convergence of ${\cal L}(\hat{\Pi},\Pi)$ for general DCMM.   
The paper presented the rate of convergence of Mixed-SCORE-X, where $X$ can be any vertex hunting algorithm that is {\it efficient}; see Definition 2.1 of \cite{MSCORE}.  Especially, the paper showed that the four variants of SVS (SVS$^*$, SVS$_0$, CVS, and SP) 
in Table~\ref{tab:CompSVS} are efficient, provided some mild regularity conditions hold (SVS$^+$ is similar to 
SVS$^*$, and can be shown to be efficient as well, under some mild regularity conditions).

For example, suppose we use SVS$_0$ or SP for vertex hunting. Write $\theta=(\theta_1,\theta_2,\ldots,\theta_n)'$ and let $\theta_{\max}$, $\theta_{\min}$ and $\bar{\theta}$ be the maximum, minimum and average of $\theta_i$'s, respectively. Additionally, write $\bar{\theta}_*=\sqrt{n^{-1}\|\theta\|^2}$. \cite{MSCORE} showed that with probability $1-o(n^{-3})$, 
\begin{equation} \label{MSrate} 
{\cal L}(\hat{\Pi},\Pi)\leq C\frac{K^3}{n\bar{\theta}^2\lambda^2_{\min}(P) }\cdot\alpha_n(\theta), \qquad\mbox{where}\quad \alpha_n(\theta)= \begin{cases}
\theta_{\max}\bar{\theta}^{3}/(\theta^2_{\min} \bar{\theta}_*^2), & \mbox{for Mixed-SCORE-SVS$_0$},\cr
\log(n) \theta^{3}_{\max}\bar{\theta}^{3}/(\theta^2_{\min} \bar{\theta}_*^4), & \mbox{for Mixed-SCORE-SP}. 
\end{cases}
\end{equation} 
Here, $\lambda_{\min}(P)$ is the minimum eigenvalue (in magnitude) of $P$, and both $(K, P)$ may depend on $n$. 
The factor $\alpha_n(\theta)$ depends on the degree configuration and the vertex hunting algorithm we use in Mixed-SCORE.  In the case of severe degree heterogeneity, the $\alpha_n(\theta)$ for Mixed-SCORE-SP can be much larger than that for Mixed-SCORE-SVS$_0$, so the convergence rate of the latter is much faster than that of the former.  This confirms our point in Section \ref{subsec:SVS}: 
directly using SP for vertex hunting may significantly underperform, and  it is desirable 
to use SVS$^*$ and SVS$^+$ instead for they have a denoise step. The theoretical result is also consistent with the numerical results in Figure~\ref{fig:VH} where we compare SP with SVS$^*$ and SVS$^+$.  
 
In the above discussion, we use the MSE ${\cal L}(\hat{\Pi},\Pi)$  to measure the error.  \cite{MSCORE} also studied the convergence rate 
when we use the maximum entry-wise $\ell^1$-norm to measure the error. Take the $\hat{\Pi}$ from Mixed-SCORE-SP for example. They showed that 
\[
\max_{1\leq i\leq n}\|\hat{\pi}_i-\pi_i\|_1\leq CK^{3/2} |\lambda_{\min}(P)|^{-1}\sqrt{\alpha_n(\theta)}, \qquad \mbox{with probability $1-o(n^{-3})$}. 
\]

For DCMM models  with moderate degree heterogeneity (i.e., $\theta_{\max}\leq C\theta_{\min}$), Mixed-SCORE is rate-optimal.  For example, \cite{jin2017sharp} presented a  minimax lower bound, where they showed that, for any estimator $\hat{\Pi}$, ${\cal L}(\hat{\Pi},\Pi)\geq c/(n\bar{\theta}^2)$ with a non-vanishing probability, where $c > 0$ is a constant. Therefore, in this case,  the rate of Mixed-SCORE-SVS$_0$ is optimal if $K$ is bounded and $|\lambda_{\min}(P)|\geq C$ (if we neglect the $\log(n)$ factor, then Mixed-SCORE-SP is also rate optimal in this case; note that rate of Mixed-SCORE-SP can be much slower if we have severe degree heterogeneity). 
For optimality of Mixed-SCORE in more general settings (e.g., severe degree heterogeneity, growing $K$, vanishing $\lambda_{\min}(P)$, etc.), see \cite{MSCORE-opt}.

\subsection{Sharp entry-wise bounds on the leading eigenvectors of the adjacency matrix $A$} 
\label{subsec:entrywise}  
As before, let $\widehat{\Xi}=[\hat{\xi}_1,\hat{\xi}_2,\ldots,\hat{\xi}_K]$ be the matrix containing the first $K$  eigenvectors of $A$, and let $\Xi=[\xi_1,\xi_2,\ldots,\xi_K]$ be the matrix containing the first $K$ eigenvectors of $\Omega$. For many spectral approaches to network analysis (e.g.,   SCORE in Section \ref{sec:SCORE} for community detection and     Mixed-SCORE in Section \ref{sec:MME} for membership estimation),  
an important technical step is to derive a 
sharp entry-wise large-deviation bound for leading eigenvectors of $A$ (i.e.,  a sharp large-deviation bound for row $i$ of $(\widehat{\Xi} - \Xi)$, for all $1 \leq i \leq n$).   
This task is much more challenging than obtaining a large-deviation bound for the sum of squared errors in all rows of $\hat{\Xi}$, because we can no longer use the classical sin-theta theorem \citep{sin-theta} to reduce it to the study of $\|A-\Omega\|$. In recent literature in network analysis, there are a few results on the entry-wise deviation bounds, mostly focusing on the more restrictive setting of $K\leq C$, $|\lambda_{\min}(P)|\geq C$ and no severe degree heterogeneity. For example, \cite{erdHos2013spectral} considered the Erdos-Renyi graph, \cite{abbe2017entrywise} considered SBM, and \cite{fan2019simple, fan2020asymptotic,liu2019community} studied network models with moderate degree heterogeneity (i.e., $\theta_{\max}\leq C\theta_{\min}$). While these results are very interesting, we can not directly use them for 
general DCMM settings here, where we may have severe degree heterogeneity. 
For general DCMM with severe degree heterogeneity,  \cite{MSCORE} derived tight entry-wise large-deviation bounds as follows. Let $\hat{r}_i$ and $r_i$ be the same as in Mixed-SCORE, $1\leq i\leq n$. They showed that, with probability $1-o(n^{-3})$, there exist two orthogonal matrices $O\in\mathbb{R}^{K,K}$ and $H\in\mathbb{R}^{K-1,K-1}$ such that 
\[
\max_{1\leq i\leq n}\|e_i'(\hat{\Xi}O-\Xi)\|\leq \frac{C\sqrt{K^{3}\theta_{\max}^{3}\|\theta\|_1\log(n)}}{\|\theta\|^3|\lambda_{\min}(P)|}, \qquad\max_{1\leq i\leq n}\|H\hat{r}_i-r_i\|\leq \frac{C\sqrt{K^3 \theta_{\max}^{3}\bar{\theta}^{3} \log(n)}}{\theta_{\min}\bar{\theta}_*^2  \sqrt{n\bar{\theta}^2}|\lambda_{\min}(P)|}. 
\]
These bounds play a key role in our study on SCORE and Mixed-SCORE, where the networks may have severe degree heterogeneity.

\section{SCORE normalization and simplex structure for topic modeling}  \label{sec:TM} 
So far, we have shown that the SCORE normalization is useful in alleviating  
the (nuisance) effect of degree heterogeneity in network analysis. 
Interestingly, for learning large-scale text data  (which can also be highly heterogeneous; see Section \ref{sec:intro}),  
the SCORE normalization can also be useful in alleviating the (nuisance) heterogeneous effect.  
Also, we find two simplex structures associated with the SCORE normalization (of course, 
despite these high-level connections, text data are very different from network data, so we need 
different models, methods, and theory).

Suppose we are given a text corpus of $n$ text documents (e.g., each of them can be the abstract of a paper) on a vocabulary of $p$ words. We summarize the data with a matrix  $D\in\mathbb{R}^{p\times n}$, where for all $1 \leq j \leq p$ and $1 \leq i \leq n$, $D(j,i)$ is the count of word $j$ in document $i$ divided by $N_i$ and $N_i$ is the length of document $i$ . The {\it probabilistic Latent Semantic Indexing (pLSI)} model \citep{hofmann1999} is a popular topic model.
Let $A_1,\ldots,A_K\in\mathbb{R}^p$ be $K$ topic vectors, where each topic vector is a Probability Mass Function (PMF) over words in the vocabulary. For each $1 \leq i \leq n$,   document $i$ is associated with a topic weight vector $w_i$, where $w_i(k)$ is the weight document $i$  puts on topic $k$, $1\leq k\leq K$ (note that $\sum_{k=1}^K w_i(k)=1$ and that $0 \leq w_i(k) \leq 1$).     
Write $D=[d_1,d_2,\ldots,d_n]$.  The pLSI model assumes that (note that in our notation, $N_i d_i$ is the $p$-dimensional  vector of word counts for document $i$) 
\[
N_id_i \sim \mbox{Multinomial}\Bigl(N_i,\;  \sum_{k=1}^K w_i(k) A_k\Bigr), \qquad 1\leq i\leq n. 
\]
Write $A=[A_1,A_2,\ldots,A_K]\in\mathbb{R}^{p\times K}$ and $W=[w_1,w_2,\ldots,w_n]\in\mathbb{R}^{K\times n}$. We call $A$ the {\it topic matrix} and $W$ the {\it topic weight matrix}. Let $\Omega = \mathbb{E}[D]$ and note that $\Omega = AW$ (so the pLSI model imposes a low-rank nonnegative factorization on $\Omega$). The problem of interest is to use $D$ to estimate $A$ and $W$. 

\subsection{The ideal simplex for the left singular vectors and the ideal simplex for the right singular vectors} 
\label{subsec:Topicsimplex}
Similar to the notion of {\it pure nodes} in DCMM, we define the {\it anchor word} \citep{donoho2003does,Ge} and {\it pure document}. 
For each $1\leq j\leq p$, define a vector $a_j\in\mathbb{R}^K$ by $a_j(k)=A(j,k)/[\sum_{\ell=1}^K A(j,\ell)]$, $1\leq k\leq K$. Word $j$ is called an {\it anchor word} if $a_j$ is degenerate (a PMF is degenerate if it is equal to one of the Euclidean basis vectors $e_1,e_2,\ldots,e_K$); furthermore, it is called an anchor word of topic $k$ if $a_j=e_k$. Document $i$ is called a {\it pure document} if $w_i$ is degenerate, and it is called a pure document of topic $k$ if $w_i=e_k$. 

SCORE can also be used in the pLSI model to produce an ideal simplex. Recall that $\Omega=\mathbb{E}[D]=AW$. Fixing any two diagonal matrices $M\in\mathbb{R}^{p\times p}$ and $H\in\mathbb{R}^{n\times n}$ that have positive diagonal entries, we consider the singular value decomposition on $M^{-1/2}\Omega H^{-1/2}$: Let $\lambda_k>0$ be the $k$th singular value, and let $\xi_k\in\mathbb{R}^p$ and $\eta_k\in\mathbb{R}^n$ be the associated left and right singular vectors, respectively. Define $R\in\mathbb{R}^{p\times (K-1)}$ and $Q\in\mathbb{R}^{n\times (K-1)}$ by
\[
R(j,k)=\xi_{k+1}(i)/\xi_1(i), \qquad Q(i,k)=\eta_{k+1}(i)/\eta_1(i), \qquad 1\leq i\leq n,1\leq j\leq p, 1\leq k\leq K-1. 
\]
Write $R=[r_1,r_2,\ldots,r_p]'$ and $Q=[q_1,q_2,\ldots,q_n]'$. The following theorems are proved by \cite{Topic}.

\begin{theorem} \label{thm:simplex-in-TM} 
{\bf (Two Ideal simplexes for pLSI}).  There is a simplex ${\cal S}_1$ in $\mathbb{R}^{K-1}$ with vertices $v_1, v_2, \ldots, v_K$, such that each row of $R$ is a convex linear combination of these vertices, $r_j=\sum_{k=1}^K\alpha_j(k)v_k$. Furthermore, $r_j$ falls on one of the vertices if  
word $j$ is an anchor word, and falls in the interior of the simplex otherwise, $1 \leq j \leq p$.  
There is a simplex ${\cal S}_2$ in $\mathbb{R}^{K-1}$ with vertices $u_1,u_2,\ldots,u_K$, such that each row of $Q$ is a convex linear combination of these vertices, $q_i=\sum_{k=1}^K \beta_i(k)u_k$. Furthermore, $q_i$ falls on one of the vertices of document $i$ is a pure document, and falls in the interior of the simplex otherwise, $1\leq i\leq n$. 
\end{theorem} 

Provided that each topic has some anchor words, we can use the rows of $R$ to retrieve the vertices of ${\cal S}_1$ and the convex combination weight vectors $\alpha_1,\alpha_2,\ldots,\alpha_p$. These weight vectors are connected to the topic matrix $A$ through the following lemmas \citep{Topic}:
\begin{lemma} \label{lemma:TM-weights} 
Write $G=M^{1/2}\mbox{diag}(\xi_1)\, [\alpha_1,\alpha_2,\ldots,\alpha_p]'\in\mathbb{R}^{p\times K}$ and let $g_1,g_2,\ldots,g_K\in\mathbb{R}^p$ denote its columns. Then, $A_k=g_k/\|g_k\|_1$, for $1\leq k\leq K$. 
\end{lemma}   
Similarly, if each topic has some pure documents, we can use the rows of $Q$ to retrieve the topic weight matrix $W$. 
The discussion is quite similar so is omitted. From a practical viewpoint, the assumption that {\it each topic has a few anchor words} is more reasonable than the assumption that {\it each topic has some pure documents}, especially for long text documents \citep{Ge, Topic, bing2020fast}.

\subsection{Estimating the topic matrix $A$ and weight matrix $W$ by Topic-SCORE}  
\label{subsec:TSCORE} 
Motivated by the simplex structure associated with the left singular vectors of $A$, \cite{Topic} proposed {\it Topic-SCORE} for estimating $(A,W)$:
\begin{itemize} \itemsep 0em
\item {\it (SCORE)}. Fix $M=\diag(D {\bf 1}_n)$ and $H=I_n$. Conduct SVD on $M^{-1/2}D H^{-1/2}$ to get the first $K$ left singular vectors $\hat{\xi}_1, \hat{\xi}_2,\ldots,\hat{\xi}_K$, $1\leq k\leq K$. Obtain the matrix $\widehat{R}\in\mathbb{R}^{p\times (K-1)}$ by $\widehat{R}(i,k)=\hat{\xi}_{k+1}(i)/\hat{\xi}_1(i)$, $1\leq i\leq p$, $1\leq k\leq K-1$. 
\item ({\it Vertex Hunting (VH)}). 
Apply a VH algorithm on rows of $\widehat{R}$ to obtain $\hat{v}_1, \hat{v}_2,\ldots,\hat{v}_K$, 
which are estimates of the vertices of ${\cal S}_1$. 
\item For each $1\leq j\leq p$, solve $\hat{\alpha}_j^*\in\mathbb{R}^K$ from the linear equations: $\hat{r}_j = \sum_{k = 1}^K \hat{\alpha}_j(k) \hat{v}_k$, $\sum_{k=1}^K\hat{\alpha}_j(k)=1$. Obtain $\hat{\alpha}_j$ by setting the negative entries of $\hat{\alpha}_j^*$ to zero and renormalize the vector to have a unit $\ell^1$-norm. Let $G=M^{1/2}\mbox{diag}(\hat{\xi}_1)\, [\hat{\alpha}_1,\hat{\alpha}_2,\ldots,\hat{\alpha}_p]'$. Normalize each column of $\hat{G}$ by the $\ell^1$-norm of this column. The resulting matrix is $\widehat{A}$.   
\item Let $\hat{W}^*=(\widehat{A}'M^{-1}\widehat{A})^{-1}\widehat{A}'M^{-1}D$. Obtain $\widehat{W}$ from $\widehat{W}^*$ setting negative entries to zero and renormalizing each column to have a unit $\ell^1$-norm. 
\end{itemize} 
Topic-SCORE was carefully analyzed by \cite{Topic}, where it was shown to be optimal in a broad class of settings; see details therein.  
One mild condition needed here is that {\it each topic has a few anchor  words}.  
Similarly, we may use the simplex structure associated with the  right singular vectors of $A$ and 
develop a different version of Topic-SCORE. The resultant procedure can be analyzed similarly, but 
we may need the condition that {\it each topic has a few pure documents}.  See  Section \ref{subsec:Topicsimplex} above for the comparison of two conditions.

\subsection{The Hofmann-Stigler's model and Topic Ranking by TR-SCORE}  
\label{subsec:TRSCORE}  
The Phase II data in Table \ref{tab:twodata} contains the abstracts of 
83,331 papers. It is of interest  to use these abstracts and the citation information in the data set  to  rank the research topics. \cite{PhaseII-AOAS} proposed the Hofmann-Stigler 
model to jointly model the citation and abstract data by combining the ideas of the  
 Stigler's model \citep{stigler1994citation} and the pLSI model above.    As above, suppose we have 
$n$ text documents (paper abstracts) satisfying the pLSI model where $w_i \in \mathbb{R}^K$ is the topic weight vector for  abstract  $i$, $1 \leq i \leq n$.   In the Hofmanm-Stigler model, we fix a topic export vector $\mu \in \mathbb{R}^K$ 
and assume that in the occurrence when paper $i$ cites paper $j$,  
topic $k$ cites topic $\ell$ for $w_i(k) w_j(\ell)$ times, where $w_i(k)$ is the weight paper $i$ puts on topic $k$; similar for $w_j(\ell)$.  
We assume that the probability that paper $i$ cites paper $j$, given that there is a citation exchange between two papers, is 
$\mathrm{exp}(\mu' (w_i - w_j)) / [1 + \mathrm{exp}(\mu' (w_i - w_j))]$.  \cite{PhaseII-AOAS} further proposed Topic-Ranking SCORE (TR-SCORE)  to rank the topics, which runs as follows. First, let $\hat{w}_i$ be the estimated topic weight above. Second, replace $w_i$ by $\hat{w}_i$ for each $i$ in the Hoffmain-Stigler's model, and obtain an estimate $\hat{\mu}$ for $\mu$ by fitting the model. Last, rank the $K$ topics using $\hat{\mu}$ (e.g., if 
$\hat{\mu}_k$ is the largest entry of $\hat{\mu}$,  then topic $k$ is ranked as the highest).

\subsection{Applications of Topic-SCORE and TR-SCORE to the analysis of the publication data of statisticians} 
\label{subsec:TPstatistics}  
As part of standard data processing,  \cite{PhaseII-AOAS} removed some  
very short abstracts and use the remaining 56,550 abstracts (from the original set of 83,331 abstracts) for analysis. 
The obtained the following results.  First, with a lot of manual efforts, \cite{PhaseII-AOAS}  identified 11 interpretable research topics in statistics (e.g.,  Bayes, Regression, Time Series) using Topic-SCORE, and obtained $\widehat{A}$ and $\widehat{W}$  for the estimates of the topic matrix $A$ the topic weight matrix $W$, respectively.  
Second, they produced a raking of the $11$ research topics using TR-SCORE, and built a knowledge graph visualizing how the ideas on one  topic influence the other topics. Last, they use the matrix $W$ to measure the research interest of an individual author and to identify the friendliest journal (among the 36 journals in the data set) for a given topic.
The matrix $\widehat{W}$ can also be used to predict whether a paper will be highly cited in the future.  \cite{PhaseII-AOAS} developed a new approach to citation prediction, where they used 
22 features, 12 of them (e.g., reference length)  being manually extracted from the data set, and 10 of them being constructed using the matrix $\widehat{W}$.  Table~\ref{tb:anchor-words} presents the 11 topics and the top $3$ anchor words (after stemming) in each topic.   See Figure 1 of \cite{PhaseII-AOAS} for details. 
 
\begin{table}[htb]
\centering
\scalebox{.82}{
\setlength\tabcolsep{3.5pt}
\begin{tabular}{l l| l  l| l  l}  
\hline
Topic & Anchor words & Topic & Anchor words & Topic & Anchor words \\
\hline
Bayes  &  paramet, nuisanc, conjug &  Hypo.Test   & stepdown, familywis, bonferroni  &  Math.Stats. &  probab, infin, ergod \\
Bio/Med.  & epidemiolog, casecontrol, alzheim &  Inference & confid, interv, coverag   &     Regression & regress, regressor, ridg  \\
Clinic.  & tnoncompli, complianc, treatment & Latent.Var. & variabl, explanatori, manifest &  Time series & time, seri, failur  \\
Exp.Design & aoptim, doptim, aberr & Mach.Learn. & scalabl, metropoli, algorithm & \\
\hline
\end{tabular}}
\caption{Top $3$ anchor words of the 11 topics estimated using the paper abstracts in the Phase II data set; see Table \ref{tab:twodata} and \cite{PhaseII-AOAS}.} \label{tb:anchor-words}
\end{table}

\subsection{Extension of Topic-SCORE and the analysis of the New York taxi data}  \label{subsec:taxi} 
State aggregation is a useful tool in optimal control and reinforcement learning. It aims to aggregate the original states of a high-dimensional Markov chain into a small number of `meta-states', in hopes of reducing the complexity of the system. \cite{duan2019state} extended Topic-SCORE to an approach for state aggregation. Let $X_0,X_1,\ldots,X_T$ be a Markov chain with $p$ states, where $p$ is presumably very large. In a {\it soft state aggregation} model, there exist latent variables $Z_0, Z_1,\ldots,Z_T\in\{1,\ldots,K\}$ such that $
\mathbb{P}(X_{t+1}=j | X_t=i ) = \sum_{k=1}^K \mathbb{P} (Z_t=k | X_t=i )\mathbb{P} (X_{t+1}=j | Z_t=k)$, for all $1\leq i,j\leq p$,  
and the distributions of $Z_t|X_t=i$ and $X_{t+1}|Z_t=k$ do not depend on $t$. Let $u_i\in\mathbb{R}^K$ be the PMF of the distribution of $Z_t|X_t=i$ and $V_k\in\mathbb{R}^p$ be the PMF of the  distribution of $X_{t+1}|Z_t=k$. Then, the probability transition matrix $\Omega\in\mathbb{R}^{p\times p}$ of the Markov chain has a decomposition: 
\[
\Omega = UV', \qquad\mbox{where}\quad U = [u_1,u_2,\ldots,u_p]'\quad\mbox{and}\quad V=[V_1, V_2, \ldots, V_K]. 
\]
We call a state $i$ an {\it anchor state for disaggregation} if exactly one of $V_1(i), \ldots, V_K(i)$ is nonzero, and an {\it anchor state for aggregation} if exactly one of $u_i(1),\ldots,u_i(K)$ is nonzero.   The soft state aggregation model imposes a topic-model-like structure on $\Omega$, where each state is treated both as a `word' and as a `document', $u_i$ and $V_k$ are analogous to $w_i$ and $A_k$ in the topic model, and the two kinds of anchor states are analogous to the definitions of anchor words and pure documents, respectively. 
The data matrix is however different. Suppose we observed multiple chains $\{X^{(m)}_t\}_{1\leq m\leq N, 1\leq t\leq T}$ and use them to obtain an empirical transition matrix $\widehat{\Omega}$, where $\widehat{\Omega}(i,j)$ is the fraction of transitions from state $i$ to state $j$ among all the chains. The interest is using $\widehat{\Omega}$ to estimate $U$ and $V$. 
\cite{duan2019state} extended Topic-SCORE to this setting, but a major difference is that the normalizing matrices $M$ and $H$ need to be chosen differently to cater to the weakly dependent noise structure in $\widehat{\Omega}$. 
\cite{duan2019state} applied this method to a New York City taxi data set. It contains the pick-up and drop-off locations of $1.1\times 10^7$ taxi trips. By discretizing the city map into $p=1922$ different locations, we can construct an empirical transition matrix $\widehat{\Omega}$ from these trips.  
They found that the estimated anchor states coincide with notable landmarks in Manhattan (see the table below); additionally, each `meta-state' (characterized by a PMF $V_k$) can be interpreted as a representative traffic mode with exclusive destinations (e.g., traffic to Times square, traffic to WTC Exchange, etc.)

\begin{table}[htb]
\hspace*{-2em}
\scalebox{.82}{
\begin{tabular}{l l| l  l| l  l| l  l}  
\hline
Meta-state & Anchor region & Meta-state & Anchor region & Meta-state & Anchor region & Meta-state & Anchor region\\
\hline
1  &  Manhattan Valley (south) &  4 & Madison Avenue  & 7 & Turtle Bay &  10 & Wall Street\\
2  & Spanish Harlem & 5  & Hell's Kitchen & 8 & Ukrainian Village &   \\
3  & Carnegie Hill & 6 & Chelsea & 9 & Tribeca & &  \\
\hline
\end{tabular}}
\caption{Interpretation of the anchor regions from the 10 estimated meta-states from the NYC taxi trip data \citep{duan2019state}.}
\end{table}


\bibliographystyle{chicago}
\bibliography{network} 


\end{document}